\documentclass[fleqn,usenatbib]{mnras}

\usepackage{newtxtext,newtxmath}
\usepackage[T1]{fontenc}

\usepackage{mathtools, cuted}

\usepackage{amsmath}
\usepackage{amsfonts}
\usepackage{amsbsy}	
\usepackage{graphicx}
\usepackage{bm}
\usepackage{color}
\usepackage{ulem}
\usepackage{commath}
\usepackage{stmaryrd}

\allowdisplaybreaks

\usepackage{hyperref}
\hypersetup{
  colorlinks=true,        % false: boxed links; true: colored links
  linkcolor=blue,         % color of internal links
  citecolor=cyan,         % color of links to bibliography
  urlcolor=cyan            % color of external links
}

\newcommand{\ie}{i.e.,~}
\newcommand{\eg}{e.g.,~}

\title{Equilibrium non-selfgravitating tori around black holes in
  parameterised spherically symmetric spacetimes}

\author[Marie Cassing]{Marie Cassing$^{1}$, Luciano Rezzolla$^{1,2,3}$\\
$^{1}$Institut f{\"u}r Theoretische Physik, Max-von-Laue-Strasse 1, 60438 Frankfurt, Germany \\
$^{2}$Frankfurt Institute for Advanced Studies,
  Ruth-Moufang-Strasse 1, 60438 Frankfurt, Germany \\
$^{3}$School of Mathematics, Trinity College, Dublin 2, Ireland
}

\date{Accepted XXX. Received YYY; in original form ZZZ}

\pubyear{2022}

\begin{document}
\label{firstpage}
\pagerange{\pageref{firstpage}--\pageref{lastpage}}
\maketitle

\begin{abstract}
Non-selfgravitating equilibrium tori orbiting around black holes have a
long history and have been employed in numerous simulations of accretion
flows onto black holes and other compact objects. We have revisited the
problem of constructing such equilibria starting from spherically
symmetric black-hole spacetimes expressed in terms of a fully generic and
rapidly converging parameterisation: the RZ metric. Within this
framework, we have extended the definitions of all of the quantities
characterising these equilibria, starting from the concept of the
von Zeipel cylinders and up to the possible ranges of the specific
angular momenta that are employed to construct families of tori.  Within
the allowed space of parameters we have then encountered both standard
``single-torus'' solutions, but also non-standard ``double-tori''
solutions. While the properties of the first ones in terms of the
presence of a single cusp, of a local pressure maximum and of a varying
outer radius, are very similar to those encountered in general
relativity, the properties of double-tori solutions are far richer and
naturally allow for configurations having the same constant specific
angular momentum and hence are potentially easier to produce in
nature. The existence of these objects is at present very hypothetical,
but if these equilibrium tori were to be observed, they would provide very
valuable information on the properties of the spacetime and on its
deviation from general relativity.
\end{abstract}

%------------------------------------------------------------------------
\section{Introduction}
\label{sec:introduction}
%------------------------------------------------------------------------

The theory of non-geodesic, perfect-fluid, non-selfgravitating,
geometrically thick and stationary tori orbiting a black hole has a long
history dating back to fundamental work in the 1970s~\citep{Fishbone76,
  Abramowicz78, Kozlowski1978}. As for any stationary fluid with compact
support in a gravitational field, its equilibrium is mainly determined by
the balance of gravitational forces, pressure gradients and centrifugal
forces. A spherical topology is natural in those configurations in which
the contributions coming from the centrifugal force are much smaller than
those due to pressure gradients and gravitational forces (\eg in a
star). On the other hand, a toroidal topology is inevitable when the
contributions to the force balance coming from the pressure gradients are
smaller than those due to the centrifugal and gravitational forces. These
are indeed the conditions of the fluid flow which we will consider
hereafter.

There are several reasons behind the important role that these
configurations have played over the years. Firstly, these configurations
are sufficiently simple that their configurations can be constructed
almost entirely analytically, thus expanding enormously our ability to
investigate radically different configurations~\citep[see,
  \eg][]{Witzany2018}. Second, because in equilibrium, these tori have
been employed for decades as initial conditions in advanced numerical
simulations of accretion flows onto black holes~\citep[see,
  \eg][]{Porth2019}, or neutron stars~\citep[see, \eg][]{Cikintoglu2022},
as they are subject to instabilities of various types when endowed with
magnetic fields~\citep{Abramowicz2011}. Third, when taken away from their
equilibrium conditions, these tori exhibit an interesting dynamics with
quasi-periodic oscillations that can be associated with those that are
observed, for instance, in high-mass X-ray binaries~\citep[see,
  \eg][]{Rezzolla_qpo_03a}. Finally, by describing the motion of the fluid
very close to the event horizon of the black hole, these tori have the
potential of providing important observational information on the
properties of the spacetime in regions of strong curvature. It is this
last aspect, in particular how the spacetime properties can be imprinted on the
characteristics of the equilibrium tori, that we will explore in
detail in this paper.

Equilibrium tori around black holes have so far been studied in
spacetimes that are either spherically symmetric or axisymmetric in
general relativity. Under these conditions, they have been shown to
appear under very generic conditions as isolated configurations or --
after a suitable work of fine tuning -- in complex nested multiple-tori
configurations~\citep{Pugliese2017}, where each torus has a
\textit{distinct} angular momentum~\citep{Pugliese2020}.
The investigations, however, have not been limited to Schwarzschild and
Kerr black-hole spacetimes~\citep[see][for an extensive
  review]{Abramowicz2011} and equilibrium tori have been studied also in
other, more exotic spacetimes. For instance, the properties -- either
equilibrium or dynamical -- of geometrically thick tori have been
explored in Schwarzschild-de-Sitter spacetimes~\citep{Rezzolla03b,
  Stuchlik2009}, in (Newman-Unti-Tamburino) NUT spacetimes
\citep{Jefremov2016}, in spherically symmetric spacetimes in
$f(R)$-gravity~\citep{Cruz2021}, around Kerr black holes with a scalar
hair~\citep{Gimeno-Soler:2019,Gimeno-Soler2021,Teodoro2021}, or more recently
in the so-called $q$-metric~\citep{Faraji2020, Memmen2021}\footnote{As a
distinct class of equilibrium solutions, electrically charged tori have
been studied in a Reissner-Nordstr\"om spacetime~\citep{Kovar2011} and in
a Reissner-Nordstr\"om-(anti-)de Sitter spacetime
\citep{Kucakova2011}.}. Finally, equilibrium tori also have been
investigated around ultra-compact objects different from black holes, such
as boson stars~\citep{Meliani2015,Olivares2020,Teodoro2020}, or naked
singularities in Kerr-de Sitter spacetimes~\citet{Stuchlik2014}.

All of these works testify both the interest in studying these fluid
configurations in general relativity and in other, alternative theories
of gravity. We here follow the same interest but focus our attention not
on a precise alternative theory of gravity or on an exotic
spacetime. Rather, our goal here is to investigate the properties of
equilibrium tori in generic and parameterised spherically symmetric
spacetimes describing either a black hole or another compact
object. While there are several options when considering parameterised
spherically symmetric spacetimes, our choice here falls on the
Rezzolla-Zhidenko (RZ) metric, which uses compactified (conformal)
coordinates and a Pad\'e-expansion in terms of continuous fractions to
achieve high accuracy already with a small number of parameters and a
straightforward treatment of the expansion. In this way, we are able to
offer a very general description of equilibrium tori around black holes
and to highlight the existence of a much richer family of tori solutions
in black-hole spacetimes that are not Schwarzschild. In particular, we
discuss the natural occurrence -- in some regions of the possible space
of parameters -- of double-tori solutions sharing the same specific
angular momentum and thus not requiring any fine tuning. Should the
existence of these tori become evident through astronomical observations,
it would provide very precise information on the properties of the
spacetime and on its deviation from general relativity.

Our work is organised as follows. In Sec.~\ref{section2} we briefly
recall the theoretical background and mathematical details necessary to
describe equilibrium tori and the von Zeipel cylinders. We 
describe the RZ-metric and its parametrization in Sec.~\ref{section3},
leaving the exploration of single-torus or double-tori solutions in
Secs.~\ref{section4} and~\ref{section5}, respectively. Finally, we
present our summary and the conclusions in Sec.~\ref{sec:conclusions}.

%------------------------------------------------------------------------
\section{The theory of equilibrium tori}
\label{section2}
%------------------------------------------------------------------------

In this Section we briefly recall the essential aspects of a
non-geodesic, perfect-fluid, non-selfgravitating, geometrically thick and
stationary torus orbiting a black hole. Since the gravitational mass of
the torus is assumed to be very small when compared with that of the
central black hole, we can exploit the test-fluid approximation and therefore
ignore the solution of the Einstein equations, relying simply on the
background metric of the given black hole, which we will employ in the
calculation of the general-relativistic hydrodynamic equations~\citep[see
  also][for more detailed discussions]{Font02a, Abramowicz2011,
  Rezzolla_book:2013}.

%-------------------------------------------------------------------------
\subsection{von Zeipel cylinders}
\label{sec:vzc}
%-------------------------------------------------------------------------

A fundamental starting point to describe such equilibrium tori is the
Newtonian theory of the von Zeipel cylinders, which will also be useful
to introduce a number of quantities that will be employed in the
remainder of this paper. Besides pressure gradients, the equilibrium in
these tori is made possible by their rotation, which we express in terms
of the angular velocity $\Omega$
\begin{equation}
\Omega:=\frac{u^\phi}{u^t}=\frac{d\phi}{dt} \,. \label{ang_velocity}
\end{equation}
and the corresponding specific angular momentum
\begin{equation}
  \ell:=- \frac{u_{\phi}}{u_t}
  \label{spec_ang_mom_l} \,.
\end{equation}
Using now from a convenient identity for the $t$-component of the four-velocity,
$ (u^{t})^{-2} = -(g_{tt}+2\Omega g_{t \phi} + \Omega^{2} g_{\phi\phi})$,
it is straightforward to obtain that
\begin{equation}
u_t^2 = \frac{g_{t\phi}^2-g_{tt}g_{\phi\phi}}{g_{\phi\phi}+2 \ell
  g_{t\phi}+\ell^2 g_{tt}} \,.
\label{ut_for_potential}
\end{equation}

After using the symmetries of the problem (\ie stationarity and
axisymmetry), the law of momentum conservation (Euler equation) can be
expressed in a very compact form as
\begin{equation}
\partial_{\mu} \ln | u_t | - \left( \frac{\Omega}{1-\Omega \ell} \right)
\partial_{\mu} \ell = - \frac{1}{\rho h} \partial_{\mu} p \,.
 \label{Euler_fluid}
\end{equation}
where $p,\rho$, and $h$ are the pressure, rest-mass density, and the
specific enthalpy, respectively. For a barotropic fluid, \ie a fluid for
which $p=p(\rho)$, the derivative of the enthalpy is proportional to the
derivative of the pressure and, as a consequence, the partial derivatives
of pressure and enthalpy commute and cancel each other
\citep{Rezzolla_book:2013}. Under these conditions, the following
identities can be proven, which constitute the thesis of the relativistic
von Zeipel theorem
\begin{equation}
  \label{eq:Omega_ell}
  \Omega  =\Omega(\ell) \,,
\end{equation}
and
\begin{equation}
\mathcal{R}^2 := \frac{\ell}{\Omega} = -
\frac{\ell(g_{\phi\phi}+g_{t\phi}\ell)}{(g_{t\phi}+g_{tt}\ell)} =
\frac{r^3\sin^2(\theta)}{r-2M} \,.
\label{Zeipel_radius}
\end{equation}
where $\mathcal{R}$ is the so-called von Zeipel (cylindrical) radius and
the last expression in Eq~\eqref{Zeipel_radius} refers to a
Schwarzschild spacetime.

Stated differently, in stationary and axisymmetric circular flow of a
barotropic fluid around a compact object, the surfaces of constant
angular velocity $\Omega$ coincide with the surfaces of constant specific
angular momentum $\ell$. Such surfaces are also known as von Zeipel
cylinders. This theorem was originally formulated in Newtonian gravity
and stated that, within a rotating selfgravitating object, {isodensity}
(or {isopycnic}, \ie at constant rest-mass density) and {isobaric} (\ie
at constant pressure) surfaces coincide if and only if the angular
velocity is a function of the distance from the rotation axis only
\citep[see, \eg][]{vonZeipel1924, Tassoul2007}. As a result, the
Newtonian von Zeipel cylinders are indeed \textit{cylinders}. In a
black-hole spacetime, however, the general-relativistic version of the
theorem, which is due to~\citet{Abramowicz71}, reveals that this is no
longer true and that the von Zeipel cylinders are cylindrical surfaces
only asymptotically~\citep{Rezzolla_book:2013}. In
Sec.~\ref{sec:tori_in_RZ} we will discuss how this theorem varies when
considering a generic and parameterised black-hole spacetime.

Another important advantage of barotropic fluids is that the differential
 $dp/ \rho h$ is an exact differential and the partial derivatives
commute $\partial_r \partial_\theta p = \partial_{\theta} \partial_r p
$. As a result, the integration of the Euler equation (\ref{Euler_fluid})
does not depend on the integration path and can be expressed as
\begin{equation}
\label{eq:Weff}
  W_{\rm eff}-W_{\rm in}= \ln | u_t | - \ln | (u_t )_{\rm in}| - \int_{\ell_{\rm
    in}}^{\ell} \left( \frac{\Omega}{1-\Omega \ell'} \right) d\ell' \,,
\end{equation}
where the index ${\rm in}$ refers to the ``inner-edge'' of the disc and $
W_{\mathrm{eff}}:=\ln |u_t|$.

%-------------------------------------------------------------------------
\subsection{General properties of geometrically thick tori}
%-------------------------------------------------------------------------

Already in Newtonian gravity, the equilibrium of a stationary rotating
fluid with compact support is determined by the balance of three factors:
gravitational forces, pressure gradients and centrifugal forces. The
relative strength of these forces will determine the geometric properties
of the fluid and, in particular, a torus topology will appear if the
contributions from pressure gradients are smaller than the contributions
from centrifugal and gravitational forces. The calculation of
equilibrium tori is particularly simple when the fluid has a constant
specific angular momentum
\begin{equation}
\label{eq:ell_const}
  \ell=\pm U={\rm const.}\,,
\end{equation}
where the $\pm$ signs in \eqref{eq:ell_const} refer to a fluid that is
either corotating ($+$) or counter-rotating ($-$) with respect to the compact
object/black hole. The advantage of tori with a constant specific angular
momentum is that, in this case, $\Omega=\Omega(g_{\mu\nu})$ [see
  Eqs. \eqref{eq:Omega_ell}--\eqref{Zeipel_radius}], that is, the fluid
angular velocity becomes an expression of the metric functions of
spacetime only, and the equipotential surfaces can be computed from the
metric coefficients and the constant specific angular momentum.

An equipotential surface that is closed at infinity, \ie $W_{\rm eff}=0$,
contains local extrema in the radial coordinate: $r_{\rm cusp}$ and
$r_{\rm max}$, which mark, respectively, the appearance of a cusp in the
effective potential and the location of the pressure (rest-mass
density) maximum. At both locations, $\partial W_{\rm eff} / \partial r =
0$, such that the corresponding specific angular momentum at these
locations is that of a Keplerian geodesic orbit. If the torus matter
fills the outermost closed equipotential surface, then $r_{\rm
  cusp}=r_{\rm in}$ represents the location on the equatorial plane where
such matter can accrete onto the black hole and where the maximum of the
effective potential on the equatorial plane is reached. Note that this
cusp is similar to the cusp appearing in a ``Roche lobe'', with the
important difference that in the latter case it corresponds to a single
point, while here to a whole circle in view of the axisymmetric nature of
these tori.  On the other hand, the minimum of the effective potential
$W_{\rm eff}$ is reached at $r_{\rm max}$ and the fluid around this location
is in a stable equilibrium, such that, if perturbed, it will oscillate
around the corresponding radial or polar epicyclic frequency
\citep{Rezzolla_qpo_03a}.

In general, the choice of the (constant) specific angular momentum will
then determine the location of the inner edge of the torus, $r_{\rm in}$,
which can be set to be between $r_{\rm cusp}$ and $r_{\rm max}$; clearly
$W_{\rm eff, in} \leq W_{\rm eff, cusp}$. The torus will also be limited
in the radial direction by $r_{\rm out}$, which is also the radial
location where $W_{\rm eff, out}=W_{\rm eff, in}$. Note that in the
region: $r_{\rm in}<r<r_{\rm max}$, the specific angular momentum is
larger than the Keplerian specific angular momentum $\ell>\ell_{_{\rm
    Kep}}$ and the orbital motion of the fluid is therefore
super-Keplerian in the inner part of the torus. On the other hand, the
opposite is true in the region $r_{\rm max}<r<r_{\rm out}$, \ie
$\ell<\ell_{_{\rm Kep}}$, such that the orbital motion of the fluid is
sub-Keplerian in the outer parts of the torus. Clearly, in these regions
the pressure gradients in the torus are needed to balance the excess
centrifugal acceleration.

\begin{figure}
  \centering
  \includegraphics[width=0.49\textwidth]{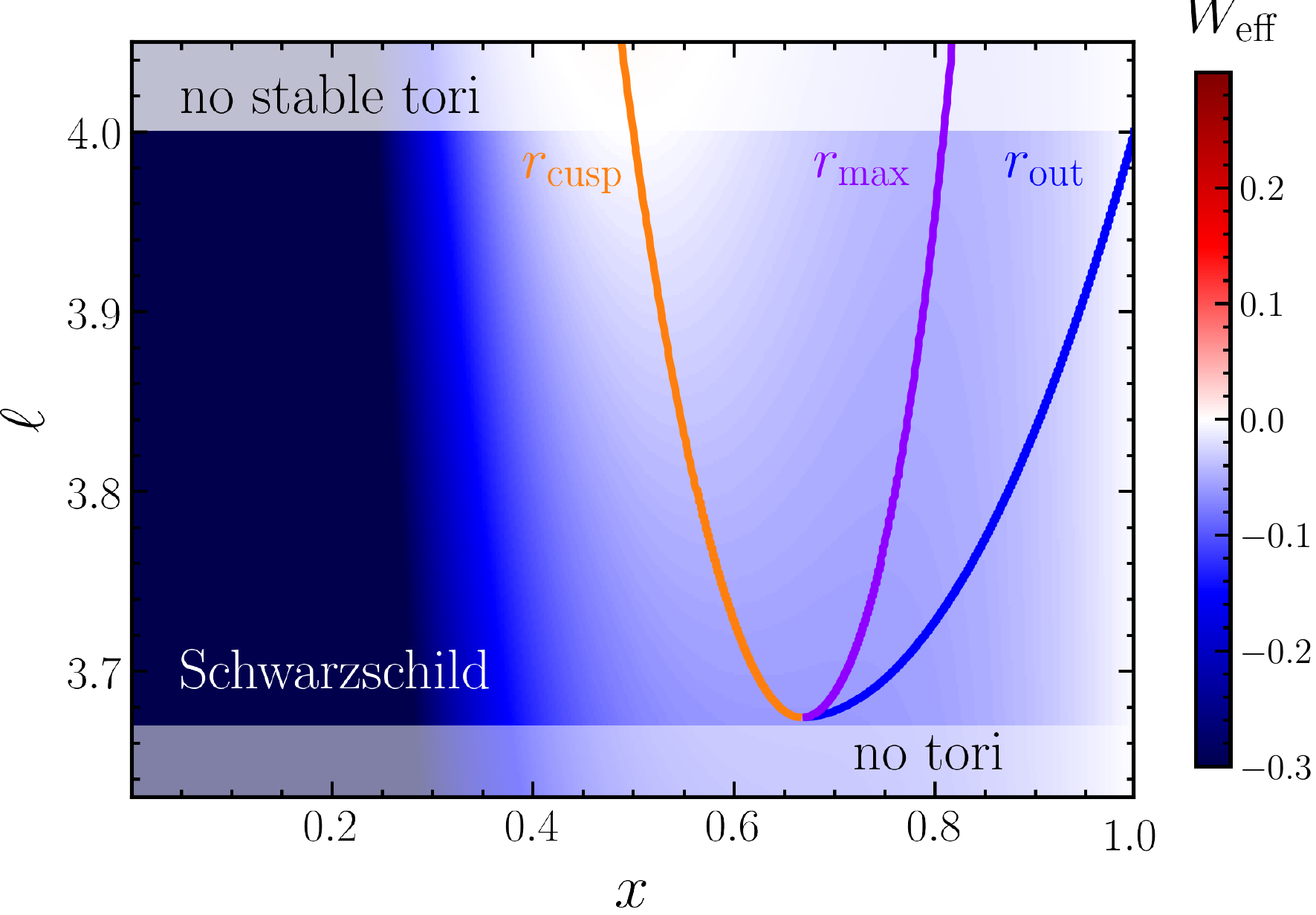}
  \caption{Colourmap of the effective potential $W_{\rm eff}$ shown as a
    function of the conformal radial coordinate $x$ and of the specific
    angular momentum $\ell$ for the Schwarzschild spacetime. The figure
    reports all of the possible tori solutions that can be built with a
    constant specific angular momentum. Shown with a solid lines are
    respectively: the location of the cusp of the torus $r_{\rm cusp}$
    (orange), of the maximum pressure $r_{\rm max}$ (purple), and of the
    outer edge of the torus $r_{\rm out}$ (blue). Transparent regions
    refer to situations in which the specific angular momentum is either
    larger than that of the marginally bound orbit, $\ell > \ell_{\rm
      mb}$ (top part) or where the specific angular momentum is smaller
    than that of the marginally stable orbit $\ell < \ell_{\rm ms}$
    (bottom part); no tori can be built in these regions.}
  \label{Slice_schwarzschild}
\end{figure}

Figure~\ref{Slice_schwarzschild} collects in a single plot the whole set
of possible tori solutions in a Schwarzschild spacetime having constant
specific angular momentum. The latter is taken to range between the
specific angular momenta of the marginally stable orbit $\ell_{\rm
  ms}=3.6738$ and of the marginally bound orbit $\ell_{\rm
  mb}=4$~\citep{Font02a,Rezzolla_book:2013}. More specifically, shown via
a colourmap is the value of the effective potential as a function of the
conformal coordinate $x$ for different values of the specific angular
momentum. Marked with different lines are the most important properties
of the torus, namely: the radial position of the cusp $r_{\rm cusp}$
(orange line), the radial position of the pressure maximum $r_{\rm max}$
(purple line), and the radial position of the outer edge of the torus
$r_{\rm out}$ (blue line). Using Fig.~\ref{Slice_schwarzschild}, which to
the best of our knowledge has not been presented before, it is
straightforward to appreciate a number of salient aspects of tori in a
Schwarzschild spacetime. First, the cusp moves towards the event horizon
as the specific angular momentum is increased. Second, in contrast to the
cusp, the position of the pressure maximum moves outwards with increasing
$\ell$. Finally, the outer radius of the torus also moves out to larger
and larger radii, reaching spatial infinity ($x=1$) for the maximum value
of the specific angular momentum. Two remarks are worth making. First,
when the specific angular momentum is larger than the one at the
marginally bound orbit, \ie for $\ell > \ell_{\rm mb}$, the tori are
effectively accreting in the sense that the outermost equipotential
surfaces will connect with the event horizon without crossing the cusp
[see, \eg, Fig. 11.8 of \cite{Rezzolla_book:2013}]. Under these
conditions, in fact, the maximum of the effective potential $W_{\rm eff}$
is larger than the corresponding value at spatial infinity; because of
this, we indicate this region with the label ``no stable tori''. Second,
if the specific angular momentum is below the value of the marginally
stable orbit of $\ell < \ell_{\rm ms}$ no equilibrium tori can be
constructed and indeed when $\ell \to \ell^{+}_{\rm ms}$, the three
fundamental scales of the torus, namely, $r_{\rm cusp},~r_{\rm max}$ and
$r_{\rm out}$ coincide; we indicate this region with the label ``no
tori''.

%------------------------------------------------------------------------
\section{Equilibrium Tori in the RZ-metric}
\label{section3}
%------------------------------------------------------------------------

We recall that the Rezzolla-Zhidenko (RZ) metric~\citep{Rezzolla2014} can
describe accurately arbitrary spherically symmetric and asymptotically
flat spacetimes representing either black-holes -- obtained in different
theories of gravity -- or other compact objects (\eg naked singularities
or boson stars)~\citep{Kocherlakota2020}. It does so by exploiting the
rapidly converging properties of a Pad\'e expansion expressed in terms of
a continuous fraction so that a small (\ie $\lesssim 4-5$) number of
parameters is sufficient to reproduce most of the spacetime metrics with
percent precision~\citep{Konoplya2020, Kocherlakota2020}. Such parameters
can be chosen such that they incorporate deviations from general
relativity and can be constrained by experimental
observations~\citep{Rezzolla2014}. The RZ parametrization solves
difficulties of other parametrizations, which normally have difficulties
in isolating the dominant terms within the corresponding parametrization,
so that a very large number of parameters is required, all with similar
strength~\citep[see, \eg][]{Cardoso2014}.  Since its development the
RZ-metric has found a number of applications \citep[see,
  \eg][]{Voelkel2020, Voelkel2021, Bauer2021, Kocherlakota2022} and some
extensions of the metric have been made in~\citep[see,
  \eg][]{Kokkotas2017,Konoplya2020b,Konoplya2020a, Bronnikov2021}. The
RZ-metric has been also used in studies of quasi-normal modes
\citep{Voelkel2019,Suvorov2021,Konoplya2022b}, and more recently, it has
played an important role in the theoretical interpretation of the
supermassive black-hole images of M87*~\citep{Kocherlakota2021} and of
Sgr A*~\citep{EHT_SgrA_PaperVI}.

In 2016, the spherically symmetric framework was extended to axisymmetric
black holes and generic compact-object stationary spacetimes
\citep{Konoplya2016a}. This was accomplished by developing two different
expansions: \textit{i)} a continued-fraction expansion in terms of a
compactified radial coordinate and \textit{ii)} a Taylor expansion in
terms of the cosine of the polar angle. Both have a fast convergence and
in the polar direction there is an exact limit on the equatorial plane.
Calculations of black holes shadow images in this axisymmetric KRZ-metric
have been performed by~\citet{Younsi2016} and the Blandford-Znajek
mechanism~\citep{Blandford1977} has been investigated
by~\citet{Konoplya2021}. A test of the KRZ-metric has been done with iron
K$\alpha$-lines by~\citet{Ni2016}, with X-ray reflection spectroscopy
by~\citep{Nampalliwar2019, Abdikamalov2021, Yu2021} or with
gravitational-wave observations~\citep{Shashank2021}. Furthermore,
\citet{Siqueira2022} have considered the KRZ parametrization to describe
a Kerr-like black hole surrounded by a massive scalar field.

In what follows, we briefly review the most salient aspects of the  RZ
approach. Using spherical polar coordinates $(t,r,\theta,\phi)$ the RZ
metric has the generic form~\citep{Rezzolla2014}
 \begin{equation}
   ds^2 = -N^2(r) dt^2 + \frac{B^2(r)}{N^2(r)}dr^2 + r^2 d\Omega^2 \,,
   \label{RZ_metric}
\end{equation}
where $d\Omega^2=d\theta^2 + \sin^2(\theta)d\phi^2$ is the line element
of a sphere. This metric is more conveniently expressed in terms of the
coordinates $(t,x,\theta,\phi)$, where $x$ is the compactified radial
coordinate defined as
\begin{equation}
  x:=1-\frac{r_0}{r} \,,
  \label{x_coordinate}
\end{equation}
 which maps the infinite interval $r\in[r_0,\infty)$ between the event
  horizon at $r_0$ and spatial infinity to the finite interval
  $x\in[0,1)$. At the horizon, the value of the metric function $N$ must
    be zero $N(r_0)=0$. This boundary condition is implicit for the
    Ansatz in the compactified coordinate $x$ as:
\begin{equation}
N^2=x A(x)  \label{N(x)} \,,
\end{equation}
where $A(x)>0$ for $0\leq x\leq 1$. After introducing the near horizon
parameters $\epsilon, a_0, b_0$ the functions $A(x)$ and $B(x)$ can be
expressed as
\begin{eqnarray}
&&A(x)=1-\epsilon(1-x)+(a_0 - \epsilon)(1-x)^2+\tilde{A}(x)(1-x^3)\,, \phantom{A(x)} \\
\label{A(x)}
&&B(x)=1+b_0(1-x) + \tilde{B}(x)(1-x)^2 \,.
\label{B(x)}
\end{eqnarray}
Here $\tilde{A}(x)$ and $\tilde{B}(x)$ characterize the near horizon
properties of the metric, \ie at $x\simeq 0$ and the properties at
spatial infinity, \ie at $x\simeq 1$. They can be parametrized in terms
of a Pad\'e-expansion as
\begin{eqnarray}
  &&\tilde{A}(x)=\cfrac{a_1}{1+\cfrac{a_2 x}{1+\cfrac{a_3 x}{1+....}}} \,,\\
\label{eq:Pe_A}
  &&\tilde{B}(x)=\cfrac{b_1}{1+\cfrac{b_2 x}{1+\cfrac{b_3 x}{1+....}}} \,,
\label{eq:Pe_B}
\end{eqnarray}
where, at the horizon, we neatly have that
\begin{equation}
\tilde{A}(0)=a_{1}\,,  \qquad \tilde{B}(0)=b_{1} \,.
\end{equation}
An important and highly effective feature of the RZ expansion is that if
any of the coefficients $a_i$ and $b_i$ with $i \geq 1$ is zero, all the
others are automatically zero, such that the truncation of the expansion is
sharp from that order on.

%-------------------------------------------------------------------------
\subsection{Equilibrium tori in the RZ-metric }
\label{sec:tori_in_RZ}
%-------------------------------------------------------------------------

Recalling the theory of the von Zeipel cylinders reviewed in
Sect.~\ref{sec:vzc}, we know that the relevant expressions to describe
the tori do not depend on the $g_{rr}$ metric function and therefore are
independent of the function $B(x)$ in the metric~\eqref{RZ_metric}. As a
result, the expression for the $t$-component of the 4-velocity is given
by
\begin{equation}
  u_t^2 = \frac{x A(x)r_0^2 \sin^2(\theta)}{-(1-x)^2 x A(x) \ell^2 + r_0^2 \sin^2(\theta)} \,,
\label{RZ_ut_for_potential}
\end{equation}
while the acceleration has the generic covariant components
\begin{eqnarray}
a_{\mu} =
\frac{1}{2}\frac{\partial_{\mu}(-N^2(x))+\Omega^{2}\partial_{\mu}\left(
  r_0^2 \sin^2(\theta)/(1-x)^2 \right)}{-N^2(x)+\Omega^{2} \left( r_0^2
  \sin^2(\theta)/(1-x)^2 \right)} \,. \nonumber \\
\end{eqnarray}

From these quantities and recalling that the acceleration must vanish for
a particle moving on a Keplerian circular orbit, it is possible to derive
the expression of the Keplerian angular velocity as
\begin{equation}
\Omega_{_{\rm Kep}} := \left(\frac{u^\phi}{u^t}\right)_{_{\rm Kep}} = \pm
\sqrt{\frac{A(x) + x A'(x)}{2 r_0^2 /(1-x)^3}} \,,
\label{RZ_omega_sol}
\end{equation}
while the corresponding expression for the Keplerian specific angular
momentum is
\begin{equation}
\ell_{_{\rm Kep}}(x) := -\left(\frac{u_\phi}{u_t}\right)_{_{\rm Kep}} =
\pm \frac{r_0}{x} \sqrt{\frac{A(x) + x A'(x)}{2(1-x)A^2(x)}} \,.
\label{spec_ang_mom_RZ}
\end{equation}

Under these conditions and a metric that is diagonal (as the RZ metric),
the equation of the von Zeipel cylinders reduces to the known relation
between $\ell$ and $\Omega$, \ie
\begin{eqnarray}
g_{tt}\ell+g_{t\phi}(1+\Omega \ell) +\Omega g_{\phi\phi}&=&0  \,, \nonumber \\
g_{tt}\ell &=& - \Omega g_{\phi\phi} \,.
\end{eqnarray}
such that the von Zeipel radius (squared) in the RZ-metric is given by
\begin{equation}
\mathcal{R}^2_{\rm vZ} := \frac{\ell}{\Omega} = -
\frac{g_{\phi\phi}}{g_{tt}}= \frac{r_0^2 \sin^2(\theta)}{(1-x)^2 N^2(x)}
\,.
\label{eq:rvz_RZ}
\end{equation}
Furthermore, the effective potential in the RZ-metric -- expressed in terms
of $x$ and $\theta$ coordinates -- reads as follows with the last expression
referring to the potential at the Keplerian angular momentum.
\begin{eqnarray}
  W_{\mathrm{eff}}(x) &=& \ln \left\vert u_t \right\vert \nonumber \\
  &=& \frac{1}{2} \ln
\left\vert \frac{x A(x)r_0^2 \sin^2(\theta)}{-(1-x)^2 x A(x) \ell^2 +
  r_0^2 \sin^2(\theta)} \right\vert \nonumber \\ &=& \frac{1}{2} \ln
\left( \frac{ x A}{1 \mp \left( (1-x)(A + x A')\right)/\left(2x(1-x)A
  \right) } \right) \,. \nonumber\\
\label{eq:Weff_RZ}
\end{eqnarray}

\subsubsection{Analytical expressions for $\epsilon$, $a_0$ and $a_1$}

Expressions \eqref{RZ_omega_sol}--\eqref{eq:Weff_RZ} are obtained in the
RZ-metric but otherwise generic, that is, not restricted to a finite set
of coefficients $\epsilon$ and $a_i$. In practice, however, we need to
truncate the infinite expansion~\eqref{eq:Pe_A} to a finite (and possibly
small) set of coefficients. In this case, setting $a_2=0$ and taking into
account the coefficients $\epsilon$, $a_0$, $a_1$ we obtain the following
expressions for the four-velocity, the angular velocity, the specific
angular momentum, the von Zeipel (squared) radius and the effective
potential
\begin{strip}
\begin{eqnarray}
u_t^2 &\!\!\!\!\!\!=&\!\!\!\!\!\! \frac{x(1-\epsilon(1-x)+(a_0-\epsilon)(1-x)^2+a_1(1-x)^3)r_0^2
  \sin^2(\theta)}{-x((1-x)^2-\epsilon(1-x)^3+(a_0-\epsilon)(1-x)^4+a_1(1-x)^5)\ell^2
  + r_0^2 \sin^2(\theta)} \, \\
\Omega_{_{\rm Kep}} &\!\!\!\!\!\!=&\!\!\!\!\!\! \pm
\sqrt{\frac{(1-x)^3\left[1+ \epsilon(-2+6x-3x^2)+a_0 (1-4x+3x^2)+a_1
      (1-6x+9x^2-4x^3) \right]}{2 r_0^2 }} \, \\
\ell_{_{\rm Kep}} &\!\!\!\!\!\!=&\!\!\!\!\!\! \pm \frac{r_0}{x} \sqrt{\frac{1+
    \epsilon(-2+6x-3x^2)+a_0 (1-4x+3x^2)+a_1 (1-6x+9x^2-4x^3)
  }{2(1-x)\left[ 1- \epsilon (1-x)+(a_0-\epsilon)(1-x)^2 + a_1 (1-x)^3
      \right]^2}}\,,\\
W_{\rm eff}(x) &\!\!\!\!\!\!=&\!\!\!\!\!\! \frac{1}{2} \ln \left(
\frac{
  x(1\!-\!\epsilon(1\!-\!x)\!+\!(a_0\!-\!\epsilon)(1\!-\!x)^2 \!+\!a_1(1\!-\!x)^3)
}{
  1\!\mp\!(1\!-\!x)
  \left[1\!-\!\epsilon(2\!-\!6x\!+\!3x^2)\!+\!a_0 (1\!-\!4x\!+\!3x^2) \!+\!a_1
    (1\!-\!6x\!+\!9x^2\!-\!4x^3)\right]\left[
    2x (1\!-\!\epsilon(1\!-\!x)\!+\!(a_0\!-\!\epsilon)(1\!-\!x)^2 \!+\!a_1(1\!-\!x)^3)\right]^{\!-\!1}
} \right) \, \nonumber \\ \\
\mathcal{R}^2_{\rm vZ} &\!\!\!\!\!\!=&\!\!\!\!\!\! \frac{r_0^2 \sin^2(\theta)}{x
  \left[(1\!-\!x)^2\!-\!\epsilon(1\!-\!x)^3\!+\!(a_0\!-\!\epsilon)(1\!-\!x)^4\!+\!a_1 (1\!-\!x)^5
    \right]} \,.
\end{eqnarray}
\end{strip}

%-------------------------------------------------------------------------
\subsection{Constraints on the range of parameters of the RZ-metric}
%-------------------------------------------------------------------------

Although the expressions presented above are given in terms of three RZ
coefficients $\epsilon, a_1$ and $a_2$ and that these coefficients are
presently unknown, it is worth noting that post-Newtonian (PN)
experiments in the solar system already provide some constraints on at
least one of them. In particular, the experiments suggest that the
parameterised post-Newtonian coefficients $\beta$ and $\gamma$ are
constrained to be~\citep{Will:2006LRR}
\begin{equation}\label{PPN}
|\beta-1|\lesssim2.3\times10^{-4},
\qquad|\gamma-1|\lesssim2.3\times10^{-5}
\,,
\end{equation}
such that
\begin{align}
 a_0 &= (\beta-\gamma)\frac{2M^2}{r_0^2} \simeq 10^{-4}\,,
\end{align}
and can be reasonably assumed to be zero in a first approximation (see
below).

For the RZ parameters considered here, analytical bounds can be derived
on the possible range of values and these have been
presented in a number of studies and recently been summarised
by~\citet{Kocherlakota2022b}. We next briefly review them also in
connection to the existence of solutions of non-selfgravitating tori. We
start by recalling that the condition for the presence of an event
horizon in the RZ-metric is $N(r_0)=0$. Outside the outermost horizon the
function $N(r)$ has to be positive and therefore $A(x)>0$. Furthermore,
the square of the specific angular momentum has to be positive, such that
using expression~\eqref{spec_ang_mom_RZ} we find that
\begin{equation}
  A(x) + x A'(x) >0  \qquad {\rm for} \quad x\in [0,1]\,.
  \label{main_cond}
\end{equation}

Evaluating the condition~\eqref{main_cond} at $x=0$ and $x=1$ we obtain
the following constraints on the parameters $\epsilon, a_0$, and $a_1$
\begin{eqnarray}
&&a_1 \geq -1 + 2 \epsilon - a_0\,, \label{cond1} \\
&&\epsilon > -1 \,. \label{cond2}
\end{eqnarray}
To find the range of allowed parameters $\epsilon$, $a_0$, $a_1$ we look
at the behaviour of the metric functions in a boundary case. For example,
when $\epsilon=0$, $a_0=0$ and $a_1\neq0$, we obtain that the largest
allowed value of $a_1$ for the specific angular momentum to be positive
is $a_1=4$.

\begin{figure*}
  \centering
  \includegraphics[width=0.33\textwidth]{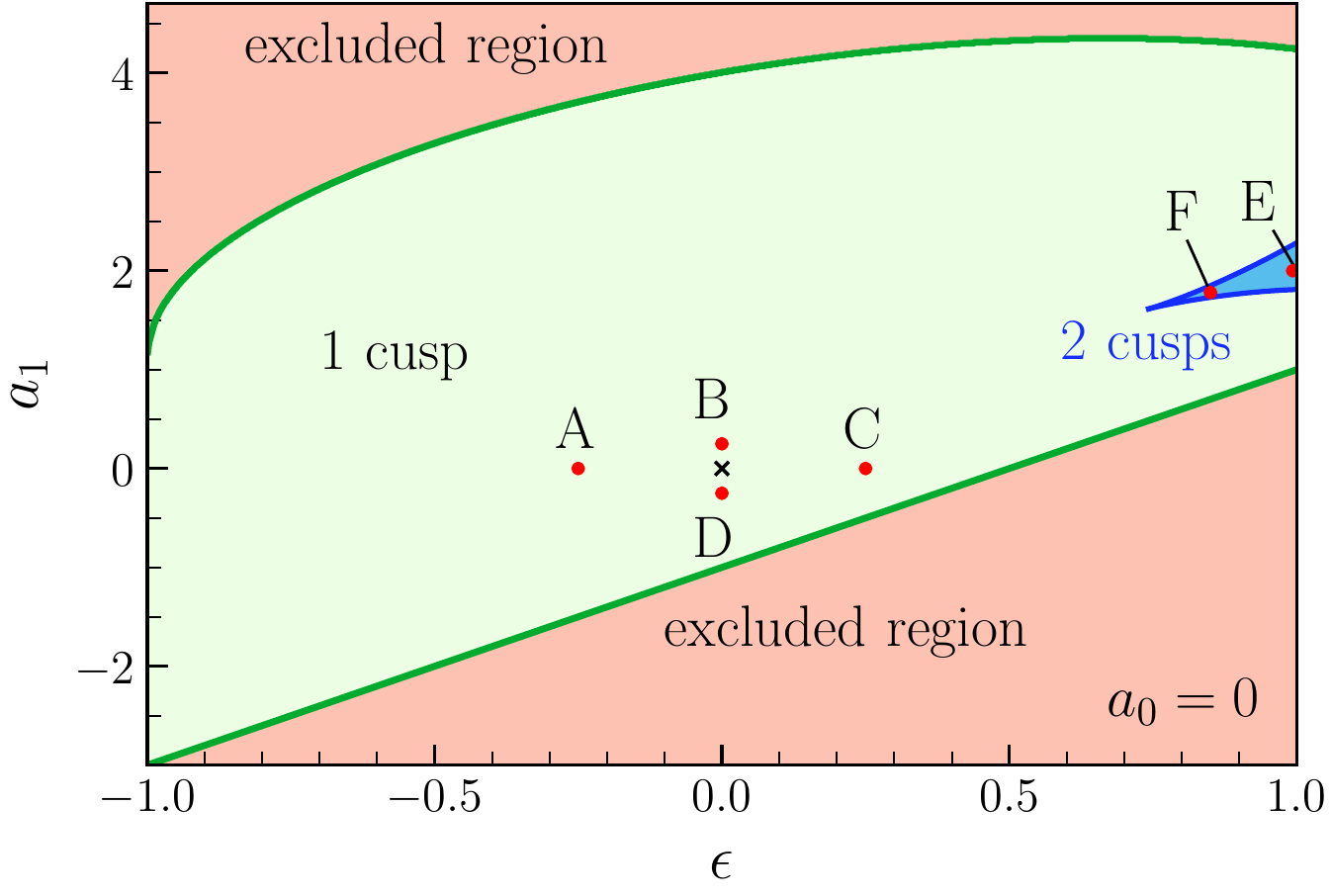}
  \includegraphics[width=0.33\textwidth]{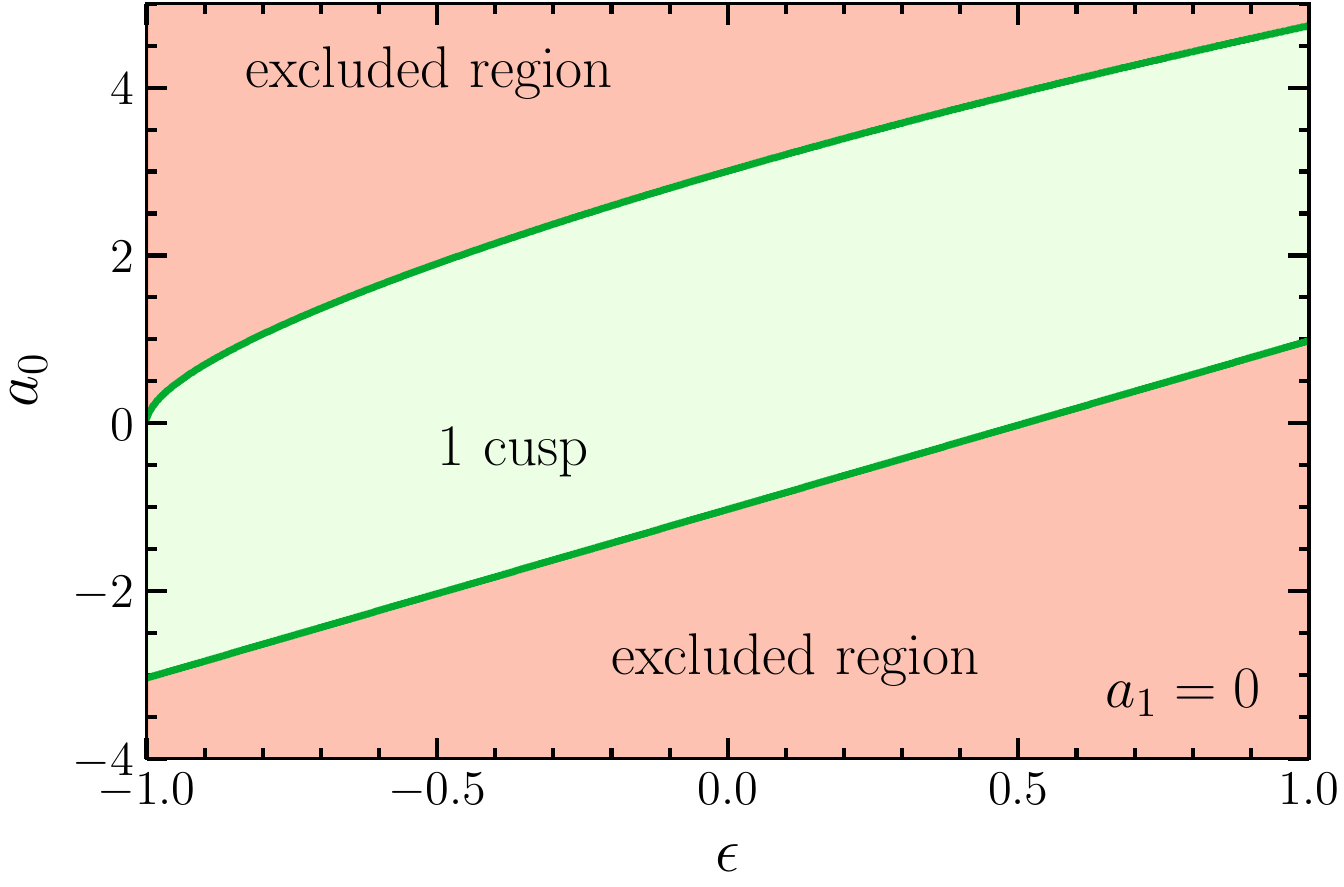}
  \includegraphics[width=0.33\textwidth]{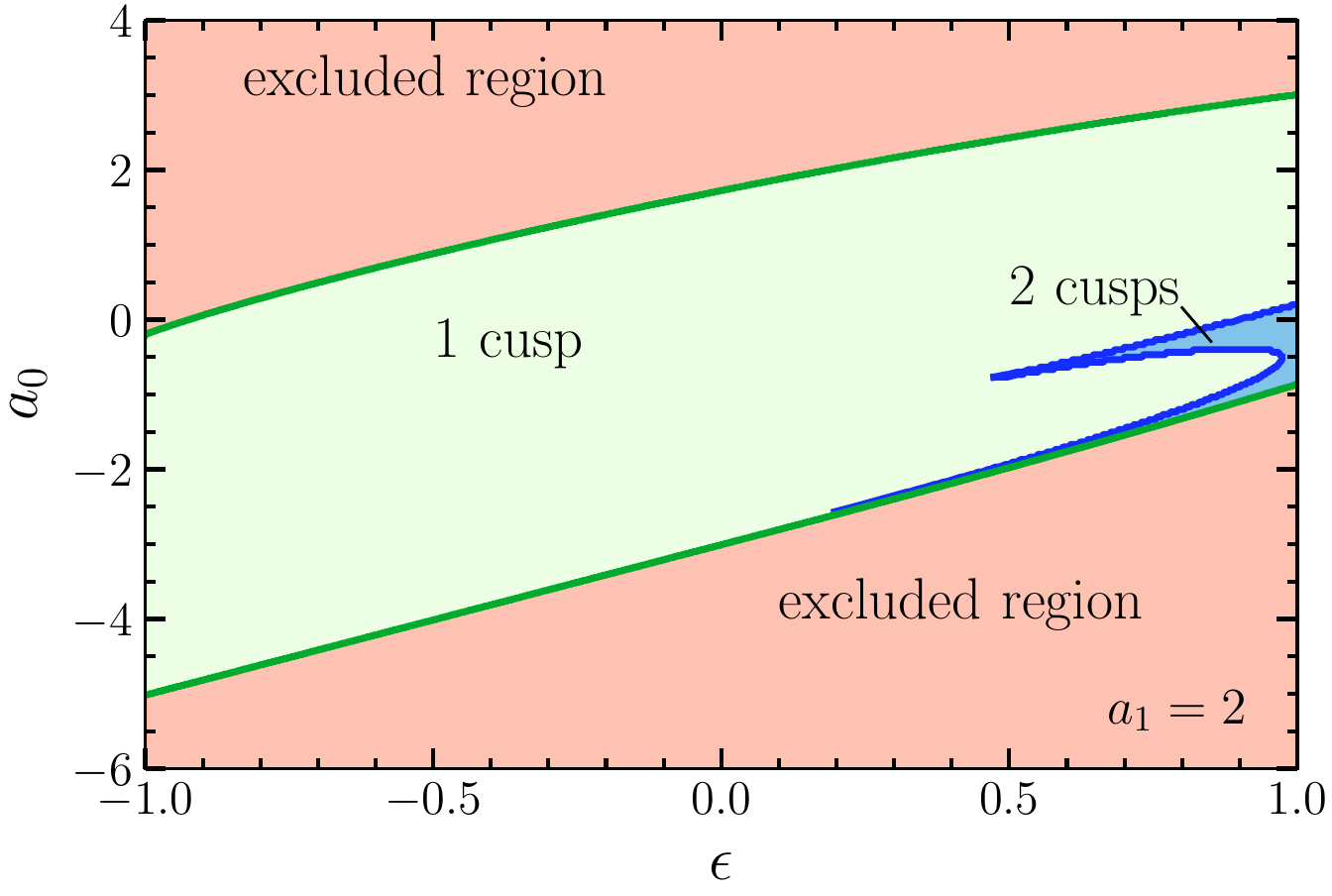}
  \caption{Multidimensional parameter space for tori solutions for the
    three RZ parameters considered here: $\epsilon, a_0$ and $a_1$. Shown
    as green-shaded are the regions where tori can be constructed and
    these are limited by solid green lines, with the lower one enforcing
    Eq.~(\ref{cond1}) and the upper one Eq.~(\ref{cond3}); no tori
    solutions can be built in the salmon-shaded regions. From left to
    right the different panels refer to sections where $a_0=0$, $a_1=0$,
    and $a_1=2$, respectively. The blue-shaded areas represent instead
    the regions where double-tori solutions are possible. Finally, red
    circles are used to mark the position of the representative models
    that are discussed in the text, while the black cross marks the
    position of a Schwarzschild spacetime.}
  \label{param_range}
\end{figure*}

To find an analytic expression for the upper boundary of $a_1$ we
consider the condition that the minimum $x_m$ of the function $A(x) +
xA'(x)$ has to be positive, \ie
\begin{equation}
  A(x_m) + x_m A'(x_m) \geq 0   \,.
  \label{cond3}
\end{equation}
The minimum is found analytically to be at
\begin{eqnarray}
  x_{m_{\pm}} &=& \left(\frac{9 a_1 - 3 \epsilon+3a_0}{12
    a_1}\right)\nonumber \\
  &&\pm
\sqrt{\left(\frac{9 a_1 + 3a_0  - 3 \epsilon}{12 a_1}\right)^2+\left(\frac{3
    \epsilon -2a_0 - 3 a_1}{6 a_1}\right)}\,.
\end{eqnarray}
For the derivation of this condition the following expressions have been
used:
\begin{eqnarray}
F(x) &=& A(x)+ x A'(x) \nonumber \\
  &=& 1+\epsilon(-2+6x-3x^2)+a_0 (1-4x+3x^2) \nonumber \\
   && +a_1(1-6x+9x^2-4x^3) \,,
\end{eqnarray}
and
\begin{eqnarray}
F'(x)= 6\epsilon(1-x) +a_0(6x-4)+a_1(-6+18x-12x^2) \, .
\end{eqnarray}
%

%------------------------------------------------------------------------
\section{Space of solutions: single tori}
\label{section4}
%------------------------------------------------------------------------

Figure~\ref{param_range} shows with different shadings the various
regions where solutions as single-torus solutions can be found (green
shading) or not (salmon shading). More specifically, the green-shaded
area -- where solutions are possible -- is upper-limited by the
condition~\eqref{spec_ang_mom_RZ} and lower-limited by the condition
~\eqref{main_cond}; in such a region, the specific angular momentum only
has a single minimum and the effective potential shows a single cusp. The
left panel in Fig.~\ref{param_range} refers to the case when $a_0=0$,
while the remaining two are computed when $a_1=0$ and $a_1=2$,
respectively.  In what follows (see Sec.~\ref{Single_a0}), we will
discuss four models in this green region: ${\rm A, B, C}$, and ${\rm D}$,
which are marked by red dots and surround the Schwarzschild solution; the
latter obviously corresponds to the case of $\epsilon =0$, $a_1 =0$ and
is marked with a black cross. Also shown in Fig.~\ref{param_range} (with a
blue-shaded area) is the region in which double-tori solutions are
possible. The specific angular momentum here has two minima and one
maximum, which allows to distinguish these solutions from the
single-torus ones. The effective potential, on the other hand, has two
cusps and two maxima, which can be filled with fluid to obtain two tori.

\begin{figure*}
  \centering
     \includegraphics[width=0.49\textwidth]{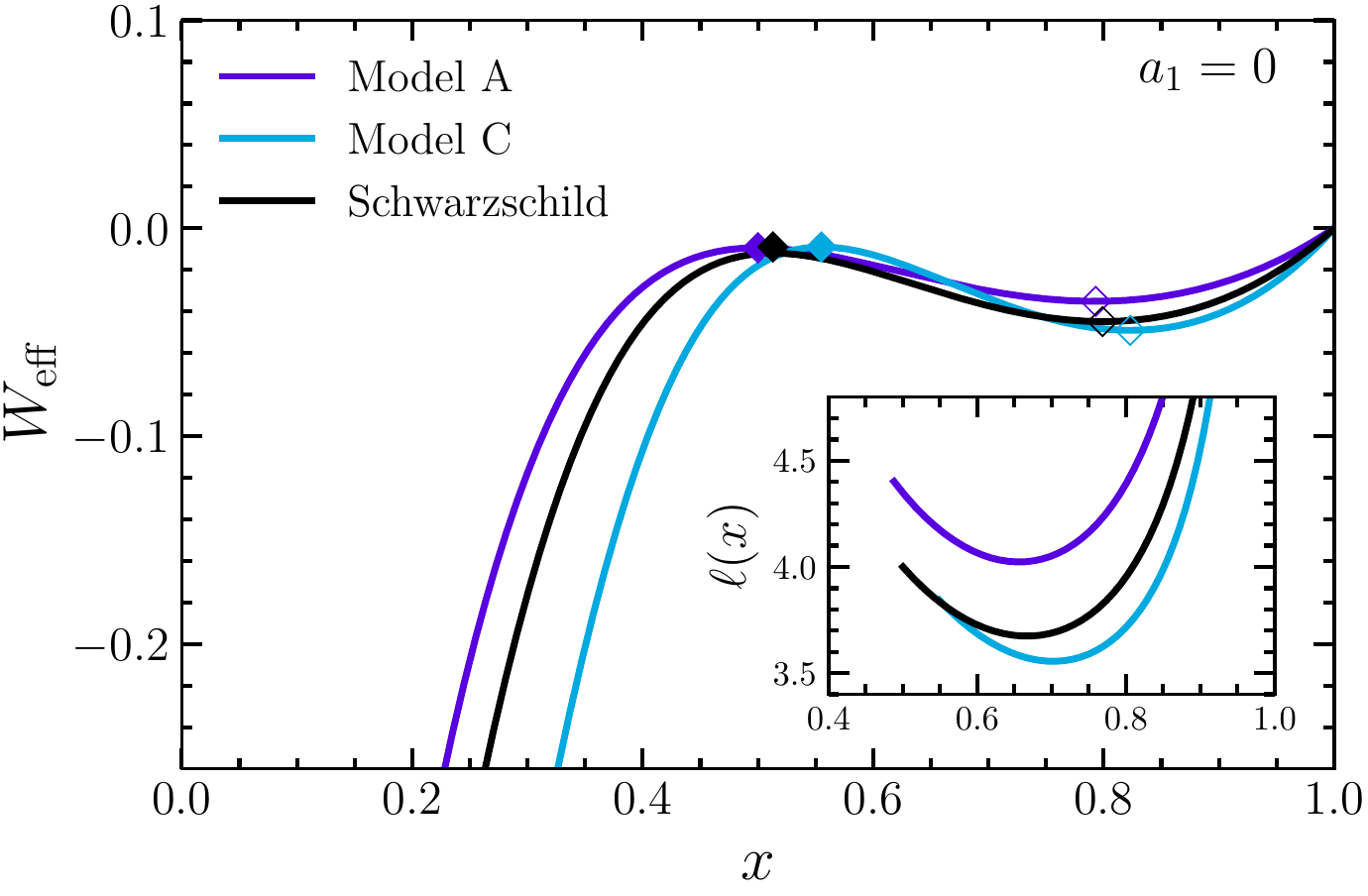}
     \includegraphics[width=0.49\textwidth]{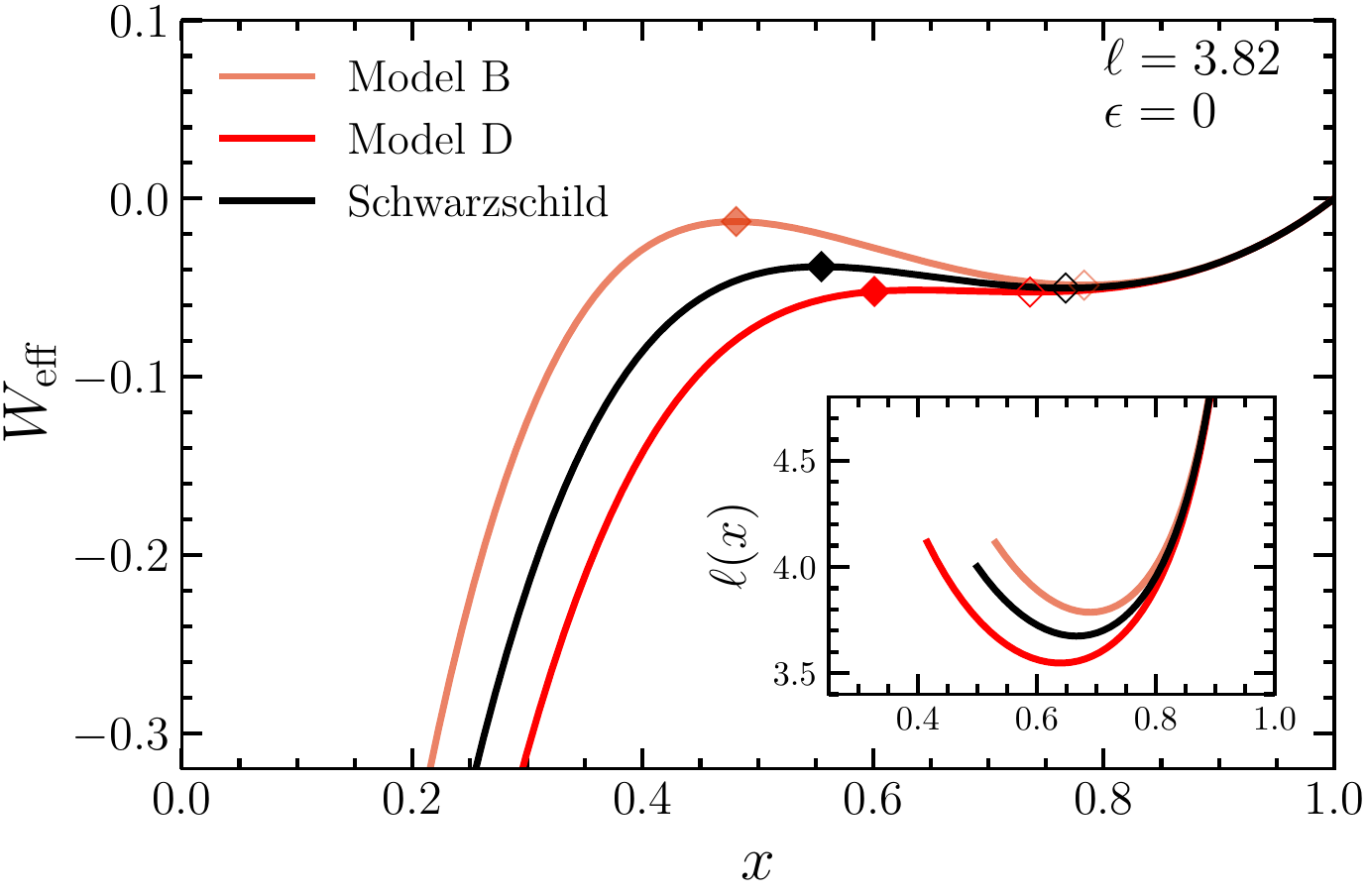}
     \caption{\textit{Left panel:} the effective potential at a fixed
       constant specific angular momentum in units of the ADM-mass $M$
       for Model ${\rm A}$ at $\ell=4.35$ (dark-blue solid line), Model
       case ${\rm C}$ (light-blue solid line) at $\ell=3.82$; also
       reported is $W_{\rm eff}$ in the case of a Schwarzschild spacetime
       with $\ell=3.9$ (black solid line). \textit{Right panel:} the same
       as in the left but for Model cases ${\rm B}$ (light-red solid
       line) and ${\rm D}$ (dark-red solid line) and the Schwarzschild
       solution at $\ell=3.82$ (black solid line). In both cases we mark
       with filled (unfilled) diamonds the position of the maximum
       (minimum) of the effective potential. Finally, shown with insets
       in both cases is the spatial dependence of the specific angular
       momentum, which is terminated at the marginally bound orbit.}
  \label{models025}
\end{figure*}

Further below (see Sec.~\ref{section5}), we will also discuss in the
double-tori region two models: ${\rm E}$, and ${\rm F}$. If the
fluid in these three models fills up the outermost equipotential surface,
then the cusp of the outer torus is connected to the outer edge of the
inner torus. Since the models in the middle and right panels of
Fig.~\ref{param_range} do not offer specific qualitative differences from
those presented in the right panel, they will not be discussed in detail
here.

%-------------------------------------------------------------------------
\subsection{Single-torus solutions ($a_0 = 0$)}
\label{Single_a0}
%-------------------------------------------------------------------------

As anticipated above, fixing the value of the specific angular momentum
allows us to investigate the changes in the potential with the parameters
$\epsilon$ and $a_1$. Hence, by suitably choosing the parameters
$\epsilon$, and $a_1$, we are able to construct four representative
models, 
\begin{align}
&{\rm ~~Model~A\!:} &\hspace{-1.5cm} \epsilon =          -0.25\,,  &&a_1 = \phantom{-}0.00\,,\nonumber\\
&{\rm ~~Model~B\!:} &\hspace{-1.5cm} \epsilon = \phantom{-}0.00\,, &&a_1 = \phantom{-}0.25\,,\nonumber\\
&{\rm ~~Model~C\!:} &\hspace{-1.5cm} \epsilon = \phantom{-}0.25\,, &&a_1 = \phantom{-}0.00\,,\nonumber\\
&{\rm ~~Model~D\!:} &\hspace{-1.5cm} \epsilon = \phantom{-}0.00\,, &&a_1 =           -0.25\,.\nonumber
\end{align}

\begin{figure*}
  \centering
  \includegraphics[width=0.49 \textwidth]{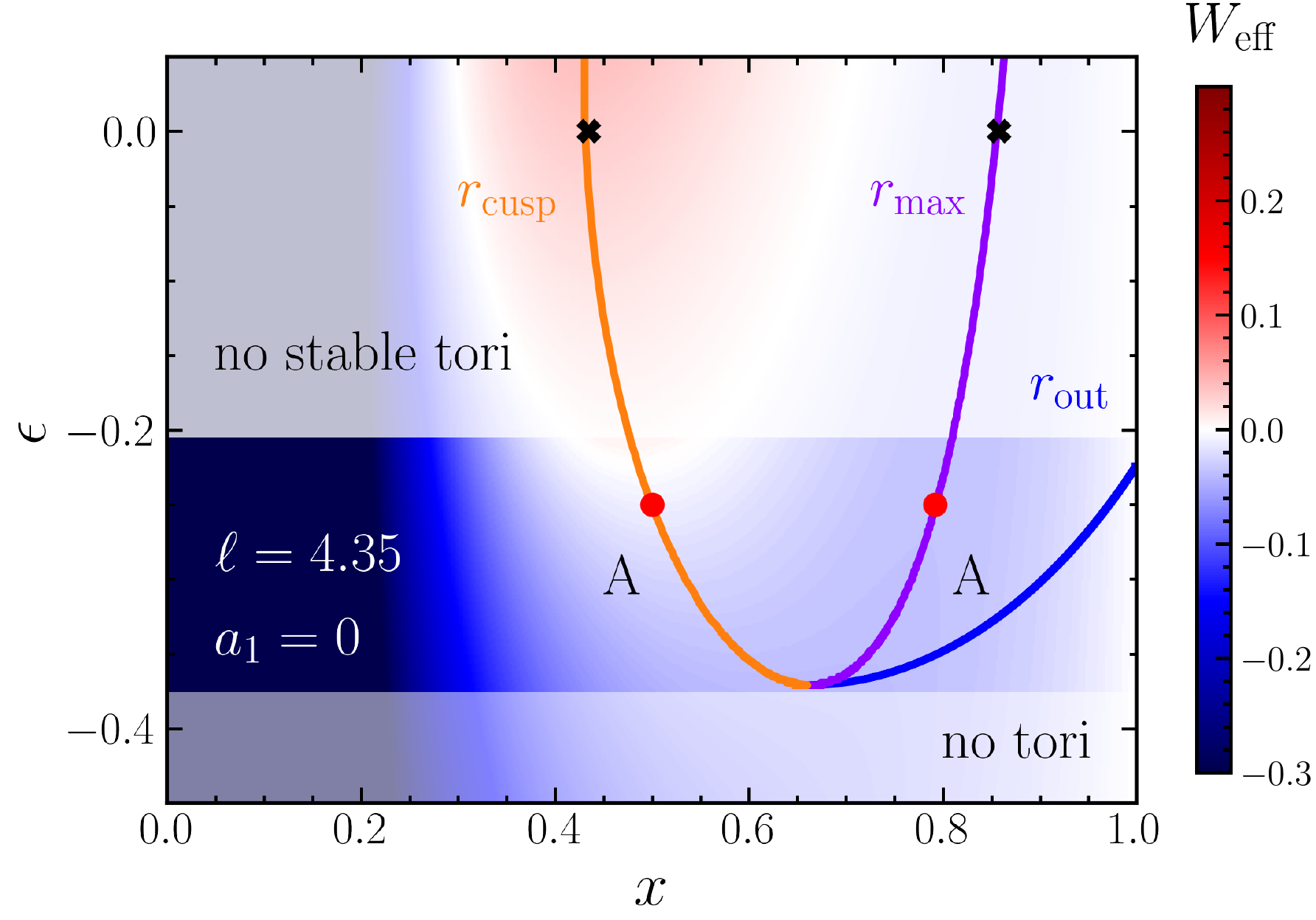}
  \hskip 0.25cm
  \includegraphics[width=0.49 \textwidth]{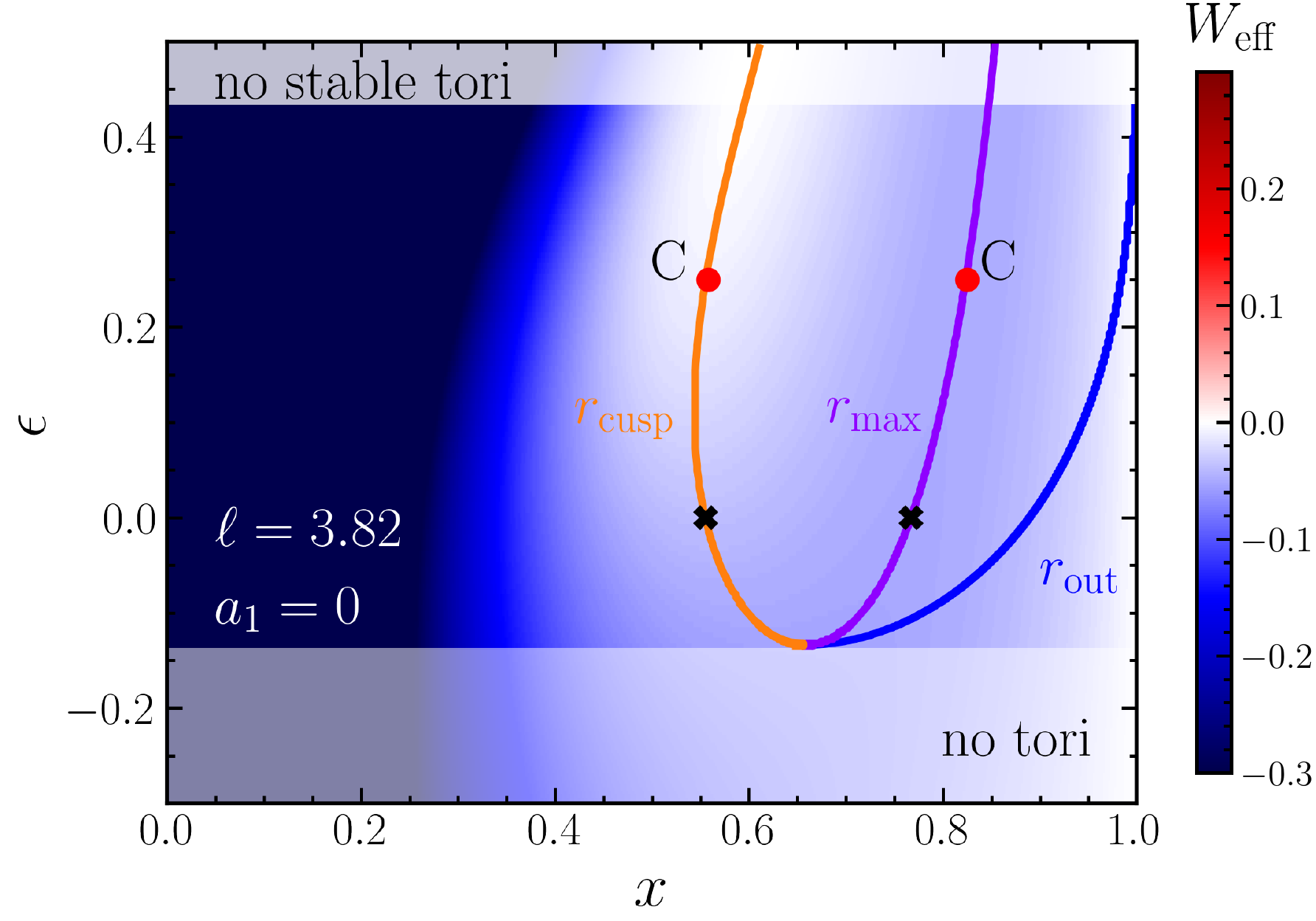}
  \caption{\textit{Left panel:} the same as in
    Fig.~\ref{Slice_schwarzschild} but for Model ${\rm A}$ in a RZ
    spacetime with parameters $a_0 = 0 = a_1$; in this case 
    the coefficient $\epsilon$ is varied and the specific angular
    momentum is kept fixed at the representative value
    $\ell=4.35$. \textit{Right panel:} the same as in the left panel but
    Model ${\rm C}$ and $\ell=3.82$.}
  \label{Slices_a1_epsi}
\end{figure*}

\begin{figure}
  \centering
  \includegraphics[width=0.49 \textwidth]{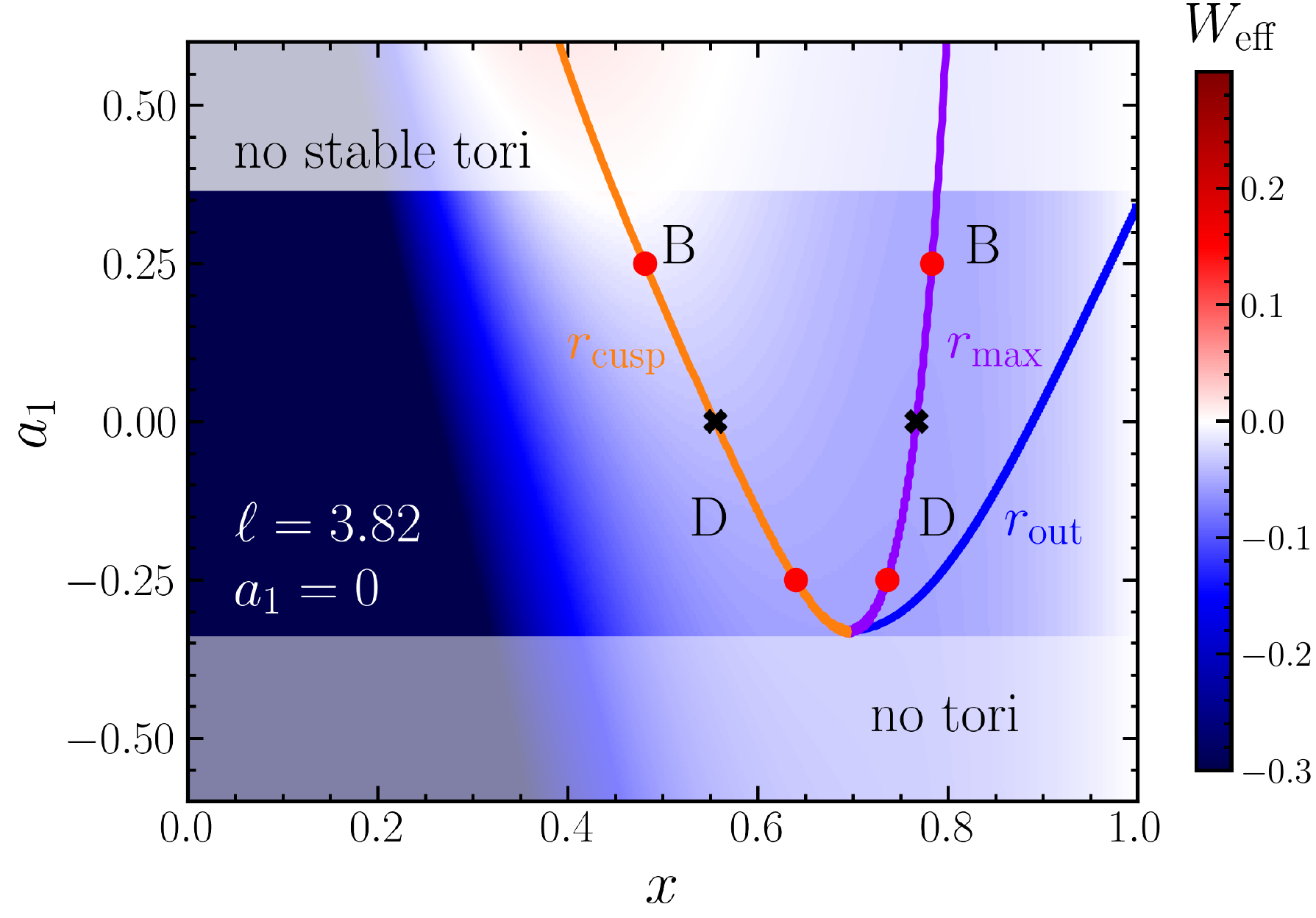}
  \caption{The same as in Fig.~\ref{Slices_a1_epsi} but for Models
    ${\rm B}$ and ${\rm D}$ in a RZ spacetime with parameters $\epsilon =
    0 = a_0$; in this case  the coefficient $a_1$ is varied and
    the specific angular momentum is kept fixed at the representative
    value $\ell=3.82$.}
  \label{Slices_a1_epsi_1}
\end{figure}

As can be seen from Fig.~\ref{models025}, the shape of the potential
changes and, in particular, the left panel contrasts Models ${\rm A}$ and
${\rm C}$ (dark and light blue lines) assessing the impact of the
parameter $\epsilon$, while the right panel contrasts Models ${\rm B}$
and ${\rm D}$ (dark and light red lines) highlighting the impact of the
parameter $a_1$. In all cases, the Schwarzschild solution is shown as a
reference by a black solid line.

\begin{figure*}
  \centering
  \includegraphics[width=1 \textwidth]{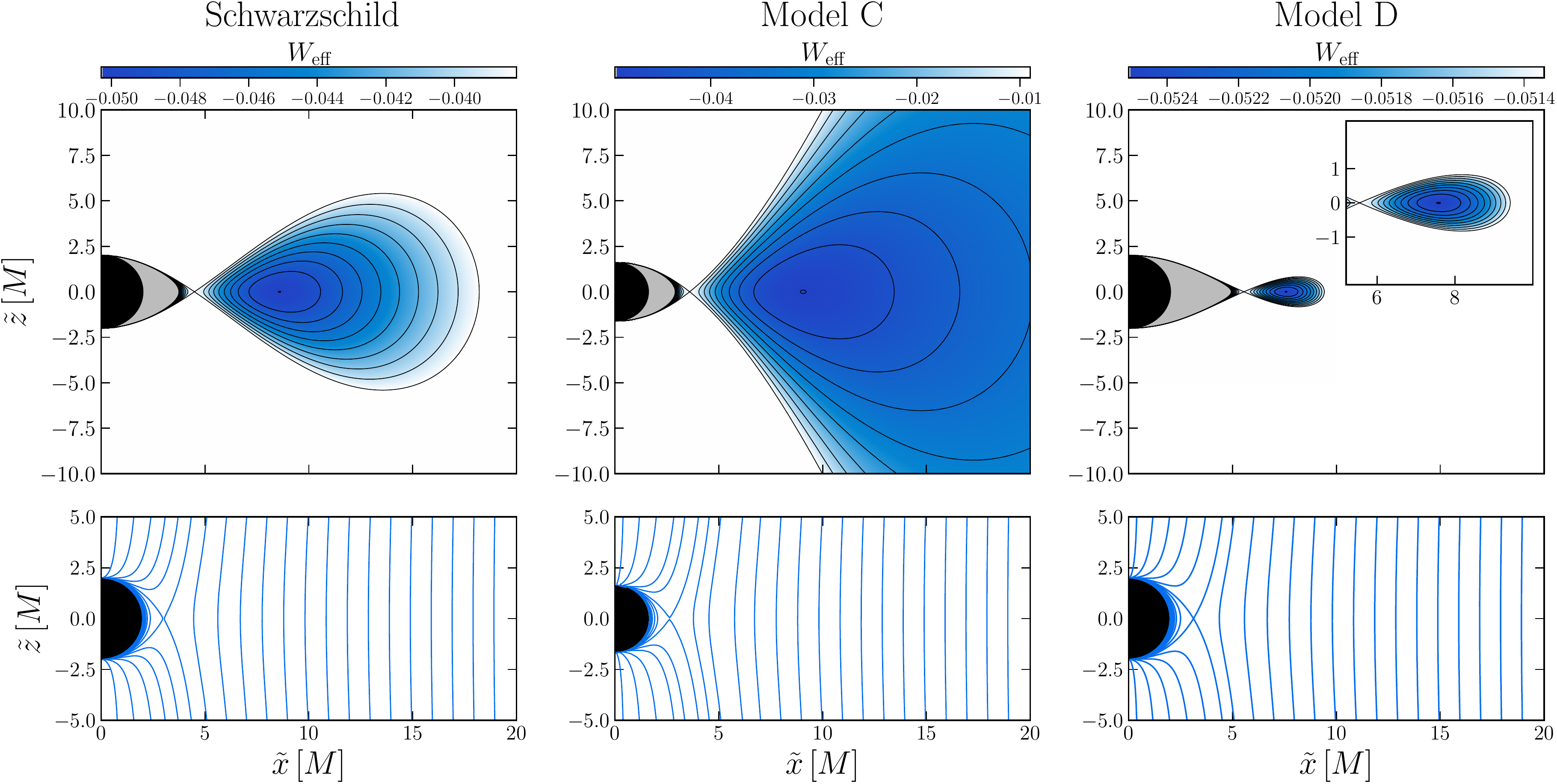}
  \caption{Equipotential surfaces of the effective potential (top row)
    and the von Zeipel cylinders (bottom row) shown in Cartesian
    coordinates ${\tilde x}$ and ${\tilde z}$ and in units of the
    black-hole mass $M$. From left to right the three columns refer to a
    Schwarzschild spacetime (left), Model ${\rm C}$ (middle), and Model
    ${\rm D}$ (right). Note that the torus in Model ${\rm C}$ has an
    outer radius $r_{\rm out}=\tilde{x}=100\,M$ and hence cannot be shown
    in the figure; furthermore, illustrated in an inset in the top-right
    panel is a magnification of the torus of Model ${\rm D}$ .}
\label{param_tori_equipot_CD}
\end{figure*}

From the study of Models ${\rm A}$ and ${\rm C}$ it is possible to deduce
that for $\epsilon < 0$ the value of the potential at the cusp and at the
maximum of the torus is smaller than the corresponding value in the
Schwarzschild spacetime (see, respectively, filled and empty diamonds in
the left panel of Fig.~\ref{models025}), since a larger value of the specific
angular momentum of $\ell=4.35$ for Model ${\rm A}$ is needed to lift the value
of the potential to the one of the Schwarzschild spacetime, which is shown for $\ell=3.9$.
At the same time for $\epsilon > 0 $  the opposite is true and the value of
the potential is larger than in the Schwarzschild spacetime needing a smaller
specific angular momentum of $\ell=3.82$ for Model ${\rm C}$. In both cases,
the position of the cusp, $r_{\rm cusp}$, moves to larger radii when considering
the same value of angular momentum.

This can be readily appreciated by inspecting the left and right panels
of Fig.~\ref{Slices_a1_epsi}, which are similar to
Fig.~\ref{Slice_schwarzschild}, and report via a colourcode the
effective potential as a function of the radial coordinate $x$ and of the
parameter $\epsilon$ at a fixed specific angular momentum $\ell=4.35$
(left) and $\ell=3,82$ (right), and RZ-parameters $a_0 = 0 =
a_1$. Clearly, the cusp is closest to the horizon for $\epsilon=0$, which
corresponds to the Schwarzschild spacetime, and then moves further away
(\ie to larger values of $x$) for increasing $|\epsilon|$ (orange
line). The left and right panels of Fig.~\ref{Slices_a1_epsi} also show
the position of the maximum-pressure of the torus $r_{\rm max}$ and
highlight that it shifts to smaller radii for negative $\epsilon$ and to
larger radii for positive $\epsilon$ as compared to the Schwarzschild
solution (purple line). In essence, therefore, for increasing
$\epsilon$, and independently of its sign, the position $r_{\rm max}$
moves to ever larger values of $x$. Similarly, the position of the outer
radius of the torus $r_{\rm out}$ (blue line) increases for
increasing values of $\epsilon$ and in the left panel it reaches spatial
infinity at $\epsilon = -0.21$, for which the value of the potential at
the cusp becomes zero. For larger values $\epsilon$ the tori would be
overfilling their Roche limit and hence no tori in stable equilibrium can
be built (see by the bright shaded region with no stable-tori solutions).

Rather similar considerations can be made for Models ${\rm B}$ and ${\rm
  D}$ (see Fig.~\ref{Slices_a1_epsi_1}), for which we
find that for $a_1 <0$ the value of the potential at the cusp is smaller
than the value of the potential in the Schwarzschild spacetime and larger
for $a_1>0$. At the same time, the cusp shifts closer to the horizon with
increasing $a_1$ (orange line), while the position and value of
the potential at the maximum of the torus changes only slightly with
$a_1$, increasing with increasing $a_1$ (purple line).
Analogous the position of $r_{\rm out}$ moves closer to spatial infinity
for increasing $a_1$.

Figures~\ref{Slices_a1_epsi} and~\ref{Slices_a1_epsi_1} also help to
appreciate via the colourcoding that the potential increases for
increasing values of the RZ parameters $\epsilon$ and $a_1$ and that a
single cusp is present for all values of $a_1$, but also disappears for
sufficiently small and negative values of $\epsilon$, \ie for $\epsilon
\leq -0.302$ for a specific angular momentum of $\ell=4.35$ and $\epsilon
\leq -0.135$ for a specific angular momentum of $\ell=3.82$ (where the
orange and the purple lines meet). Below these values the effective
potential shows no cusp and therefore no tori solutions are possible,
which is shown by the brighter shaded region. While this behaviour is
generic, the value of $\epsilon$ at which this happens depends on the
constant specific angular momentum considered.  Finally, marked with red
circles in Figs.~\ref{Slices_a1_epsi},~\ref{Slices_a1_epsi_1} are the
four representative Models ${\rm A}$--${\rm D}$ and by simply considering
the distance in the conformal coordinate $x$ between the these points it
is simple to appreciate that the torus shape can vary considerably when
changing the RZ parameters $\epsilon$ and $a_1$ and that, as $a_1$ is
increased, the torus cusp and center systematically move-in and out,
respectively.

The very different aspects of the tori in the various cases can also be
seen from Fig.~\ref{param_tori_equipot_CD} where we present the
equipotential surfaces and the von Zeipel cylinders for the case of the 
Schwarzschild spacetime (left panel) and Models ${\rm C}$ (middle panel) and ${\rm
  D}$ (right panel). The equipotential surfaces are constructed for the
same value of the specific angular momentum of $\ell=3.82$ and are
displayed in terms of the Cartesian coordinate $\tilde{x}$ (not to be
confused with the conformal coordinate $x$) and $\tilde{z}$. Compared to
the Schwarzschild case we clearly observe that we obtain a much larger
torus for Model ${\rm C}$ and a much smaller torus for Model ${\rm
  D}$. This figure underlines the statement that the size of the torus
increases with increasing parameters $\epsilon$ and $a_1$. Model ${\rm
  C}$ has the RZ-parameter $\epsilon=0.25$ and the torus size increases
(especially in the $\tilde{z}$-direction) when compared to the
$\epsilon=0$ Schwarzschild case. In contrast, Model ${\rm D}$, has a
negative $a_1$ RZ-parameter, the torus is smaller and also less extended
vertically. Note that the von Zeipel cylinders have a cusp at different
von Zeipel radius $\mathcal{R}_{\rm vZ}$. In the Schwarzschild case, in
fact, $\mathcal{R}^2_{\rm vZ} =5.1975$, while it is $\mathcal{R}^2_{\rm
  vZ} =4.8351$ for Model ${\rm C}$ and $\mathcal{R}^2_{\rm vZ} =5.3879$
for Model ${\rm D}$.

%------------------------------------------------------------------------
\section{Space of solutions: double tori}
\label{section5}
%------------------------------------------------------------------------

We next discuss the region of the RZ space of parameters where the
specific angular momentum has two minima and the effective potential
therefore shows two cusps and two maxima for the \textit{same} constant
value of the specific angular momentum. These properties of the effective
potential, which \textit{are not} encountered in a Schwarzschild
spacetime, lead to what we refer to as ``double-tori'' solutions.

Within this region, which is shown as blue-shaded in
Fig.~\ref{param_range}, we consider again two representative model cases
obtained after setting $a_0=0$ and suitably selecting the parameters
$\epsilon$ and $a_1$, \ie
\begin{align}
&{\rm ~~Model~E\!:} &\hspace{-1.5cm} \epsilon = 1.00\,,  &&a_1 = 2.00\,,\nonumber\\
&{\rm ~~Model~F\!:} &\hspace{-1.5cm} \epsilon = 0.85\,,  &&a_1 = 1.78\,.
\end{align}

\begin{figure*}
  \includegraphics[width=0.7\textwidth]{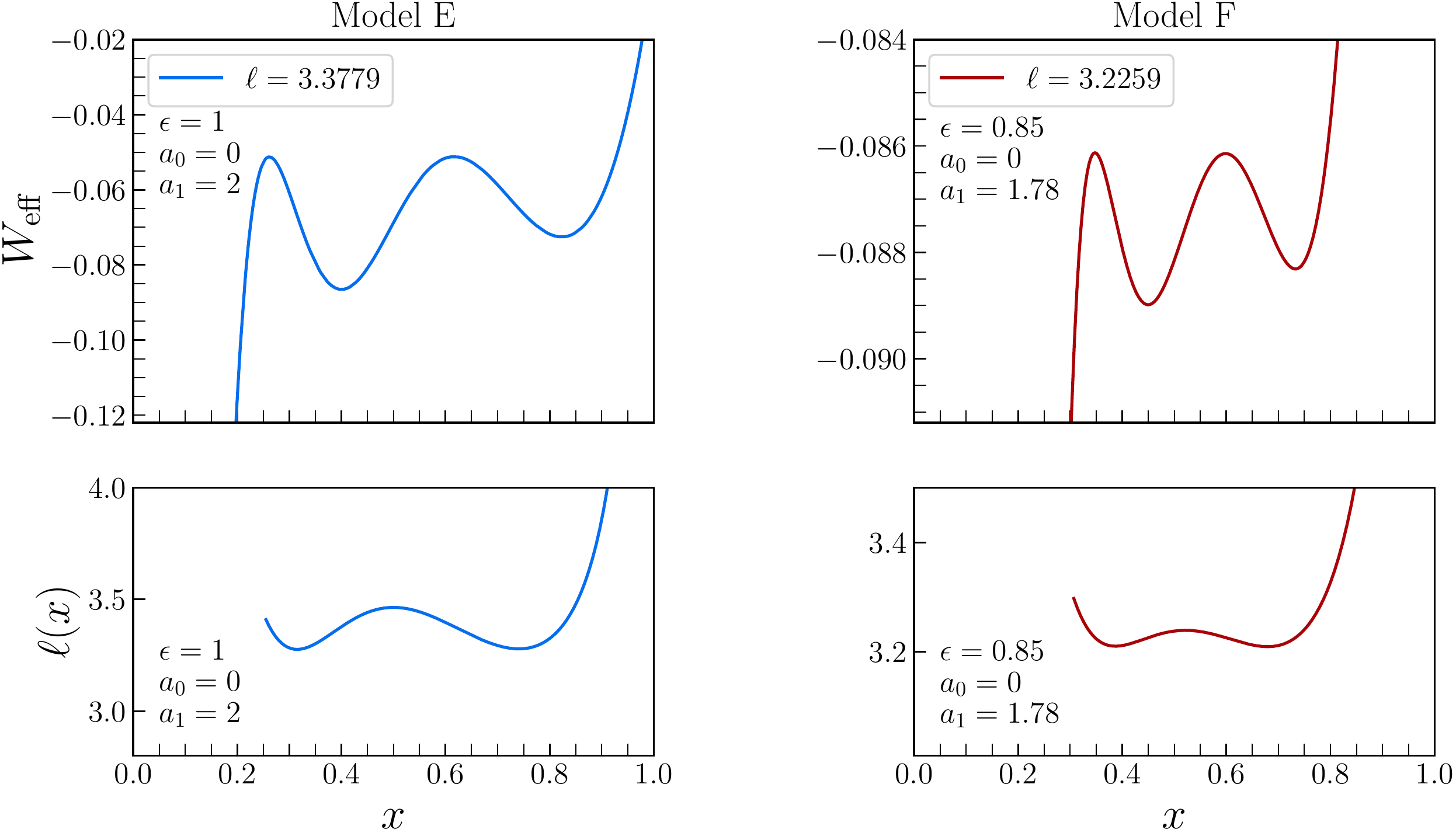}
	\caption{Effective potentials at a fixed specific angular
          momentum (top row) and spatial dependence of the specific
          angular momenta (bottom row) for Models ${\rm E}$ (left) and
          ${\rm F}$ (right). The effective potentials refer to
          $\ell=3.3779$ for Model ${\rm E}$ and to $\ell=3.2259$ for
          Model ${\rm F}$; these values are chosen so that the two cusps
          have the same value of $W_{\rm eff}$.}
	\label{Double_tori_W_L}
\end{figure*}

We should remark that it is in principle possible to build double-tori
solutions in a Schwarzschild spacetime and indeed, interesting works
exist where these configurations are considered in great detail and where
families of nested tori configurations are considered~\citep[see,
  \eg][]{Pugliese2017, Pugliese2020}. However, in these cases one needs
to suitably choose and fine-tune the specific angular momenta -- which
are \textit{different} for each torus -- such that the configuration can
actually be built. At the same time, double-tori solutions with the same
constant specific angular momentum have been found also in another
spherically symmetric black-hole spacetime that is different from the 
Schwarzschild spacetime, \ie the $q$-metric~\citep{Memmen2021}.

\begin{figure*}
  \centering
  \includegraphics[width=0.49\textwidth]{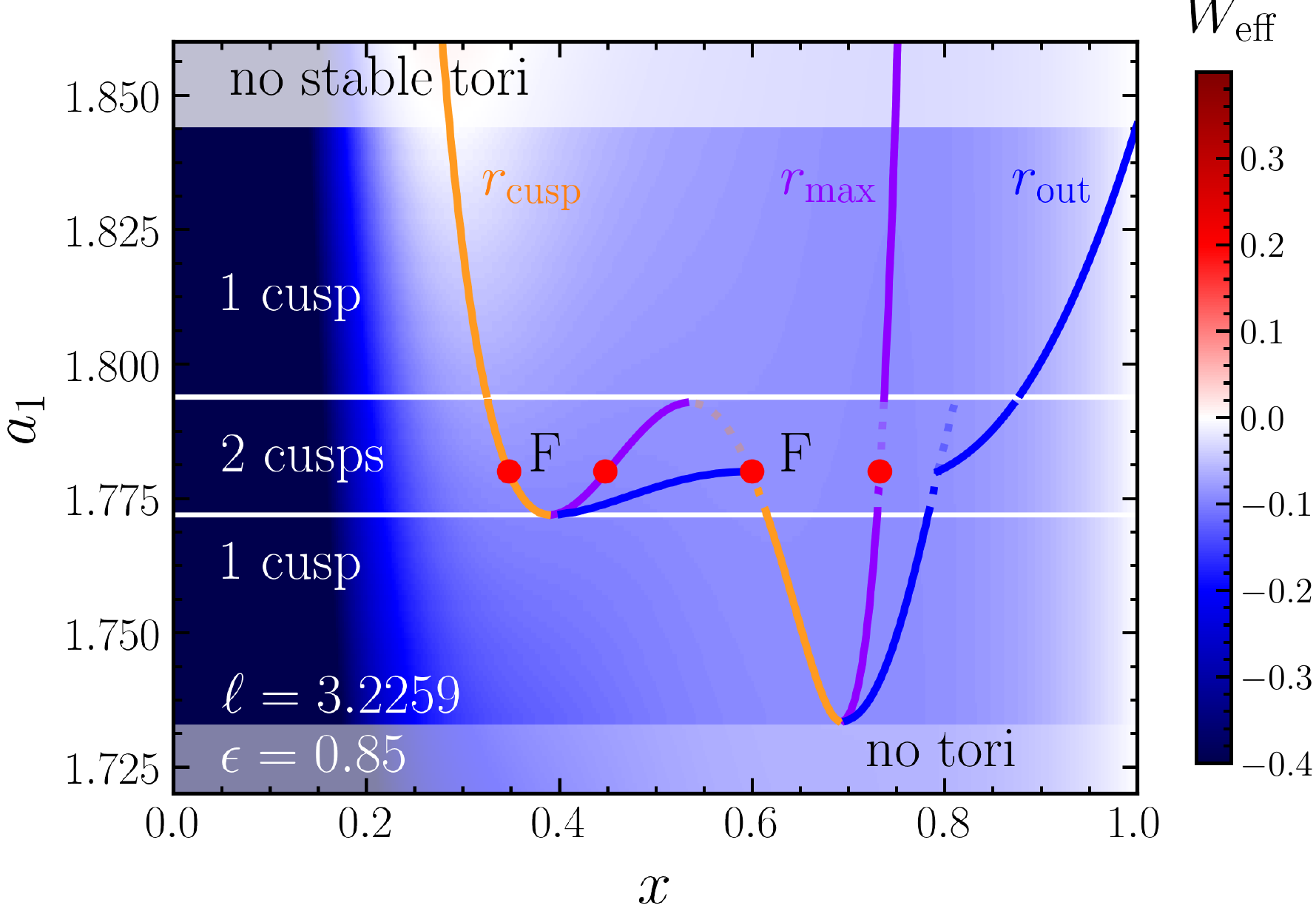}
  \hskip 0.25cm
  \includegraphics[width=0.49\textwidth]{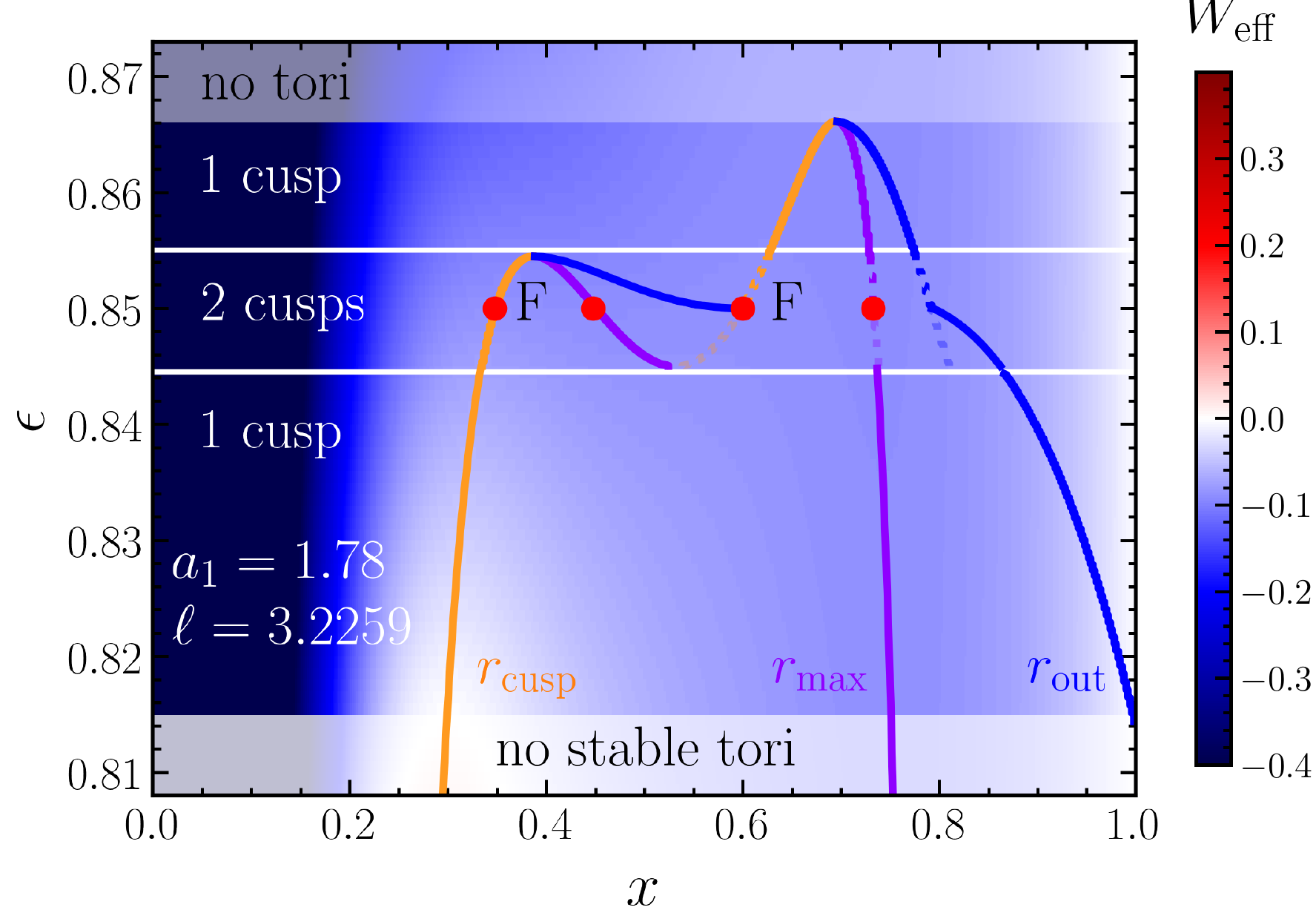}
  \caption{\textit{Left panel:} the same as in Fig.~\ref{Slices_a1_epsi}
    but for Model ${\rm F}$ in a RZ spacetime with parameters
    $\epsilon=0.85$, $a_0=0$, $a_1=1.78$; in this case the
    coefficient $a_1$ is varied and the specific angular momentum is kept fixed at
    the representative value $\ell=3.2259$. The convention for the
    various lines is the same as in the previous similar representations of
    the effective potential with the addition that we show by dotted
    transparent lines the situations in which one or two tori can exist
    depending on how the equipotential surfaces are filled. Note that in
    this case double-tori solutions are possible. \textit{Right panel:}
    the same as in the left panel but where the RZ-parameter $\epsilon$ is
    allowed to vary at fixed $a_1=1.78$.}
  \label{Slice_a1_modelF}
\end{figure*}

%-------------------------------------------------------------------------
\subsection{Double-tori solutions ($a_0 = 0$)}
\label{Double_a0}
%-------------------------------------------------------------------------

When considering Model ${\rm E}$, we note that with $\epsilon=2.00$ the
RZ parameter $a_1$ needs to be in the range $1.812 \leq a_1 \leq 2.28$
so that two well-defined cusps can appear in the effective potential,
hence our choice of $a_1=2$. The bottom-left part of
Fig.~\ref{Double_tori_W_L} shows the spatial dependence of the specific
angular momentum, which clearly exhibits two minima, while the top-left
of the same figure shows the effective potential at a fixed angular
momentum; the latter, $\ell=3.3779$, has been chosen such that the value
of the effective potential at the two cusps is the same. With increasing
values of the specific angular momentum, the potential of the inner cusp
increases and the potential of the outer cusp decreases; the opposite
happens if the specific angular momentum is decreased. Similarly, when
considering Model ${\rm F}$, we note that for $\epsilon=0.85$ the allowed
range for the RZ parameter is $1.74 \leq a_1 \leq 1.85$ so that two
well-defined cusps appear in the effective potential, hence our choice of
a representative case with $a_1=1.78$. Also in this case we display,
respectively, in the bottom-right and top-right part of
Fig.~\ref{Double_tori_W_L} the spatial dependence of the specific angular
momentum and of the effective potential for a fixed value $\ell=
3.2259$. Also in this case, and as we will discuss below more in detail,
different choices of the constant angular momentum can lead to
considerable changes of the effective potential and hence to the
appearance (or disappearance) of the outermost cusp.

\begin{figure*}
  \centering
  \includegraphics[width=0.7 \textwidth]{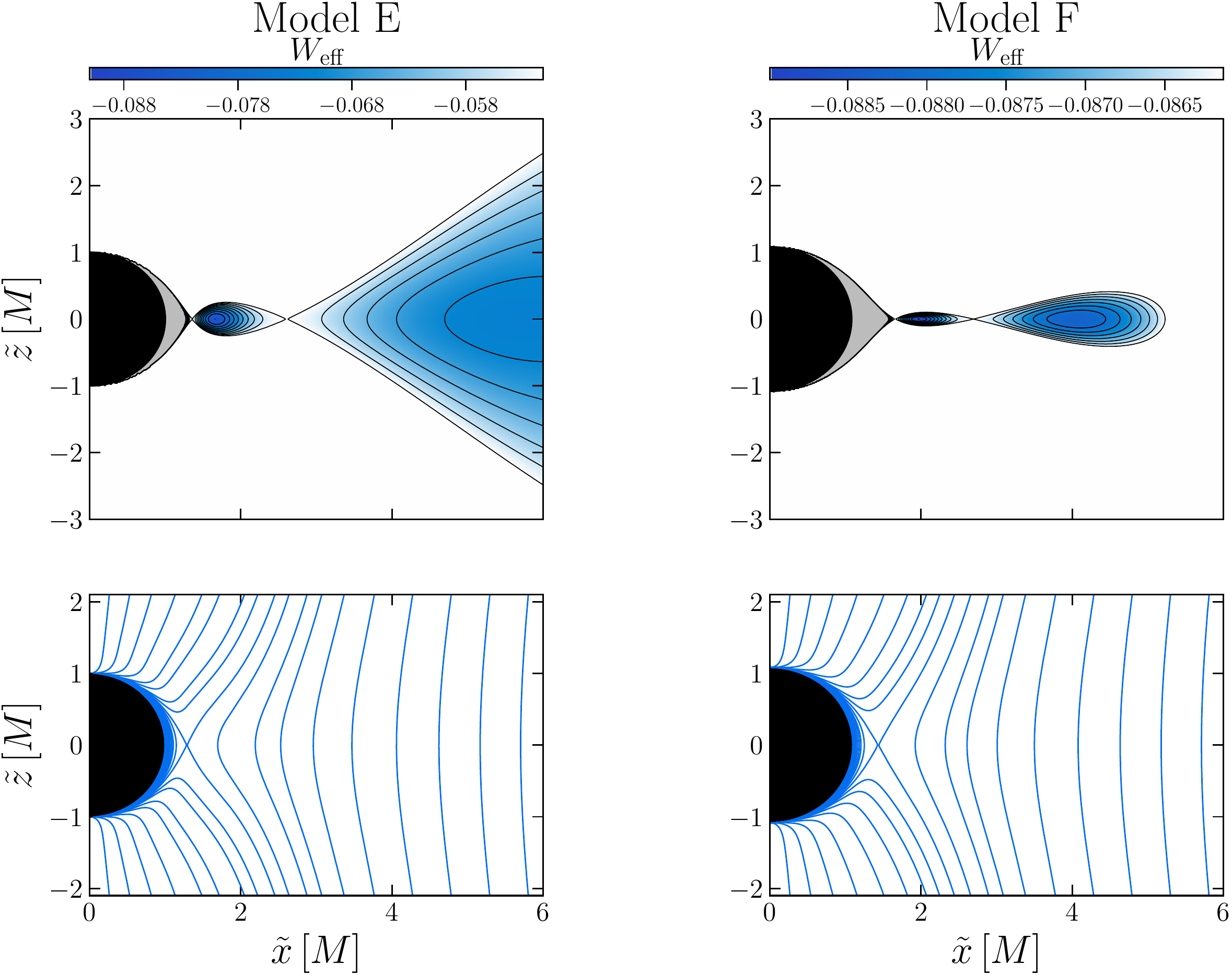}
  \caption{The same as in Fig.~\ref{param_tori_equipot_CD} but for Model
    ${\rm E}$ (left) and Model ${\rm F}$ (right). Note that in both cases
    two tori are possible and that in the case of Model ${\rm E}$ the
    outer torus is much larger than the inner one (the outer radius at
    $r_{\rm out}=\tilde{x}=13.793\,M$) and hence is
    shown only close to the second cusp.}
\label{param_tori_equipot}
\end{figure*}

Figs.~\ref{Slices_a1_epsi} and~\ref{Slices_a1_epsi_1} have already
illustrated how the changes in the constant specific angular momentum
impact on the shape of the effective potential and, in turn, on the
properties of the tori. This behaviour becomes made more complex when
double-tori solutions are possible and this is demonstrated in
Fig.~\ref{Slice_a1_modelF}, where we concentrate on the representative
Model ${\rm F}$ and again show (with a colourcode) the effective
potential for fixed values of $\ell$ and $\epsilon$ but varying values of
$a_1$, while the right panel shows the effective potential for fixed
values of $\ell$ and $a_1$ but varying values of $\epsilon$.

In analogy to Figs.~\ref{Slices_a1_epsi} and~\ref{Slices_a1_epsi_1},
marked with a transparent map in the left panel of
Fig.~\ref{Slice_a1_modelF}, are those regions where tori solutions are not
possible.  More specifically, the transparent region at low values of
$a_1$ denotes the range where the specific angular momentum is too small for the
effective potential to show a cusp and hence no tori can be constructed
there. On the other hand, in the transparent region at high $a_1$ the
tori are unstable because the chosen specific angular momentum is larger
than the value at the marginally bound orbit $\ell > \ell_{\rm mb}$.
Note that the effective potential has only one cusp for
$1.734<a_1<1.773$, such that only single-torus solutions are possible in
this range of $a_1$; we refer to this region as to the ``1-cusp region'',
where the positions of the cusps, maxima of the tori and outer radii are
presented by solid lines (orange, purple, and blue, respectively) in the
1 cusp region. In the range $1.773<a_1<a_{1,{\rm F}}$, where $a_{1,{\rm
    F}}:= 1.78$ is the value of $a_1$ for Model ${\rm F}$, a second inner
cusp and a second maximum appear in the effective potential at smaller
values of $x$. In this region, which we refer to as the ``2-cusps''
region, we mark with solid lines the relevant positions of the innermost
(first) torus and with dotted lines of the corresponding colour the
various positions of the outermost (second) torus.

In this region, both tori fill their cusps but as the value of $a_1$ is
increased to reach the value of Model ${\rm F}$ (red dots) the cusps are
at the same value of potential for the selected value of $\ell$. As a
result, for $a_1 > a_{1, {\rm F}}$ the potential of the inner cusp
increases while the one of the outer cusp decreases. This results in the
inner cusp having a value of the effective potential that is larger than
the outer cusp and we can no longer obtain two tori filling their
cusps. Hence, in the region $a_{1,{\rm F}}<a_1<1.794$ there are two
possibilities to construct tori. The first one is to fill the effective
potential up to the inner cusp (solid orange, purple and blue lines),
which yields a position of the outer radius larger than the position of
the second (outer) cusp and a single extended torus. The second
possibility is to fill the effective potential only up to the second
(outer) cusp, where the small outer torus (transparent dotted lines)
would fill its Roche lobe, but the inner one would not and therefore is
not shown. On the other hand, for $a_1>1.794$, the outer cusp disappears,
while the inner (and now only) cusp continues to move to smaller
radii, leading to single-torus solutions, which are very extended and
again represented by solid lines. Finally, and as mentioned already,
the specific angular momentum for parameters $a_1>1.844$ (transparent
region) is larger than the marginally bound orbit and the single tori in
this region are not in a stable equilibrium and will be subject to
accretion.

The right panel of Fig.~\ref{Slice_a1_modelF}, which displays the
effective potential at a fixed value of $a_1=1.78$, but with varying
$\epsilon$ is very similar to the left panel and, indeed, the
same considerations apply when going from high to low values of
$\epsilon$. More specifically, no cusp exists for $\epsilon>0.865$
(transparent top region), while a single cusp and  maximum (orange and purple
solid lines) appear in the region $0.855 < a_1 < 0.865$ giving rise to
single tori.

Two tori filling their cusp are obtained for parameters $0.85 <\epsilon
<0.855$ with the inner torus represented by solid lines and the outer torus
by dotted lines. Model ${\rm F}$ (marked by red circles) shows two cusps
at the same value of the effective potential, while for $0.845
<\epsilon<\epsilon_{F}$ the effective potential of the inner cusp rises
above the outer one and it is possible to either fill the inner cusp to
obtain one large torus (solid lines) or the outer one giving a small
outer torus (transparent dotted lines) and an inner one not filling its
cusp (not shown). With decreasing value of $\epsilon$ the outer cusp
disappears at the white solid line and we are left with a single large torus
(solid orange, purple and blue lines) in the region $0.8147 < \epsilon
<0.845$. Finally, for even smaller values $\epsilon<0.8147$ (transparent
shading) no tori in stable equilibrium can be obtained since the specific
angular momentum is above that of the marginally bound orbit.

Fig.~\ref{param_tori_equipot} provides a more direct representation of
the properties of the tori in Models ${\rm E}$ (left column) and ${\rm
  F}$ (right column) by showing, in analogy to
Fig.~\ref{param_tori_equipot_CD}, the equipotential surfaces (top row)
and the von Zeipel cylinders (bottom row). Since Models ${\rm E}$(left column)
and ${\rm F}$ (right column) differ essentially in the values of the parameter
$\epsilon$, Fig.~\ref{param_tori_equipot} shows that for an increasing
value of $ \epsilon$ also the size of the tori increases. Note also the
outer torus is systematically larger than the inner one and grows
(shrinks) faster with increasing (decreasing) $\epsilon$. Overall, the
examples discussed in this section highlight that double-tori solutions
can be easily obtained in  RZ spacetime with $a_0=0$ and in a suitably
chosen range of values for $\epsilon$ and $a_1$, which determine their
properties in terms of vertical thickness and size. More importantly,
these tori can be filled with fluids having the same constant
specific angular momentum and hence are potentially easier to produce in
nature as one expects that the matter sourcing these tori near
supermassive black holes has the same astrophysical origin and hence the
same specific angular momentum.  Double-tori solutions are also possible
to be constructed in general relativity but, in contrast, require
\textit{different} and suitably tuned values of the specific angular
momentum for each torus~\citep{Pugliese2017}.

As a final remark we note that while the effective potential in the case
of double-tori solutions has two distinct cusps, the von Zeipel radius
can only have one cusp and this has to necessarily appear close to the black
hole and not at large distances (see Fig.~\ref{param_tori_equipot}). This
is because $W_{\rm eff}$ depends on both the $g_{tt}$ metric function and
on the specific angular momentum [see Eq.~\eqref{eq:Weff_RZ}]; while the
former does not have extrema at large distances from the black hole, the
latter does, as shown for instance in Fig.~\ref{Double_tori_W_L}. By
contrast, the von Zeipel radius depends only on the $g_{tt}$ metric
function and thus has a only one local minimum at the photon ring, while
approaching asymptotic flatness at large distances [see
  Eq.~\eqref{eq:rvz_RZ}].

%------------------------------------------------------------------------
\section{Conclusions}
\label{sec:conclusions}
%------------------------------------------------------------------------

The study of non-selfgravitating equilibrium tori orbiting around black
holes has a long history and these models have found a number of
applications in the simulation of accretion flows onto black holes and
other compact objects. We have revisited the problem of constructing such
equilibria starting from the simplest black-hole spacetimes, \ie
spherically symmetric, but expressed in terms of a fully generic and
rapidly converging parameterisation: the RZ metric.

Already in general relativity, the construction of such equilibria does
not depend on the $g_{rr}$ metric function and thus our analysis has been
restricted to the first three parameters of the $g_{tt}$ metric function
within the RZ metric, \ie $\epsilon$, $a_0$ and $a_1$. Within this
framework we have extended the definitions of all of the quantities
characterising these equilibria, starting from the concept of the
von Zeipel cylinders and up to the possible ranges of the specific
angular momenta that are employed to construct families of tori.

In this way, we were able to set precise constraints on the ranges of the
RZ coefficients and thus define the space of parameters in which tori
solutions are possible and those where no such solutions can be
constructed. To make our analysis more tractable, we have further assumed
that $a_0=0$ (as in general relativity) since this is the coefficient
that is best constrained by parameterised post-Newtonian measurements to
be $a_0\lesssim 10^{-4}$. Within the allowed space of parameters we have
then encountered both standard ``single-torus'' solutions, but also
non-standard ``double-tori'' solutions. While the properties of the first
ones in terms of the presence of a single cusp, of a local pressure
maximum and of a varying outer radius, are very similar to those
encountered in general relativity, the properties of double-tori
solutions are far richer. In particular, depending on the specific region
of the space of parameters, it is possible to construct tori that have
two cusps and fill the corresponding equipotential surfaces either up to
the value of the effective potential at the first cusp or at that of the
second cusp. In this way, transitions from single-torus to double-tori
solutions (and vice-versa) are possible and the RZ parameterisation opens
therefore the way to the exploration of a much richer class of equilibria
than  in general relativity. More importantly,
these tori can be filled with fluids having the \textit{same} constant
specific angular momentum and hence are potentially easier to produce in
nature as one expects that the matter sourcing these tori close to
supermassive black holes has the same astrophysical origin and hence the
same specific angular momentum.

A concluding remark should be dedicated to the plausibility of these
solutions. Of course, all present observations of black holes, either via
gravitational waves~\citep{Abbott2016c} or via imaging
\citep{Akiyama2019_L1, EHT_SgrA_PaperI} portray a picture that is
perfectly compatible with general relativity, which still represents the
best theory of gravity presently available. At the same time,
observations are sufficiently uncertain to leave room for alternative
theories and for deviations from general relativity. If cast in this
context, and when supported by precise astronomical observations, the
existence of these equilibrium tori would provide very valuable
information on the properties of the spacetime and on its deviation from
general relativity.

%------------------------------------------------------------------------
\section*{Acknowledgments}
%------------------------------------------------------------------------

It is a pleasure to thank A. Cruz-Osorio, C. Ecker, and P. Kocherlakota
for useful discussions and comments. Support in funding comes from the
State of Hesse within the Research Cluster ELEMENTS (Project ID
500/10.006), from the ERC Advanced Grant ``JETSET: Launching, propagation
and emission of relativistic jets from binary mergers and across mass
scales'' (Grant No. 884631).

%------------------------------------------------------------------------
\section*{Data Availability}
%------------------------------------------------------------------------

Data available on request. The data underlying this article will be
shared on reasonable request to the corresponding author.

\bibliographystyle{mnras}
\bibliography{aeireferences}
%\bibliography{tori_in_RZ.bbl}

\bsp	% typesetting comment
\label{lastpage}
\end{document}